\documentclass[a4paper,11pt]{article}
\usepackage{jcappub} 
\usepackage{lineno}
\usepackage{adjustbox}
\usepackage{graphicx}
\usepackage{orcidlink}
\usepackage{aas_macros}
\usepackage[caption=false]{subfig}
\arxivnumber{2404.16401} 
\title{\boldmath Model independent approach for calculating galaxy rotation curves for low $S/N$ MaNGA galaxies}

\author[a, b]{{Sangwoo Park}\orcidlink{0009-0000-2266-9985},}
\author[a, b]{{Arman Shafieloo}\orcidlink{0000-0001-6815-0337},}
\author[c,d]{{Satadru Bag}\orcidlink{0000-0003-0141-606X},}
\author[e]{{Mikhail Denissenya}\orcidlink{0000-0003-4734-7127},}
\author[f,g]{{Eric V.~Linder}\orcidlink{0000-0001-5536-9241},}
\author[h]{{and Adarsh Ranjan}\orcidlink{0000-0001-9882-1576}}

\affiliation[a]{Korea Astronomy and Space Science Institute (KASI), 776 Daedeok-daero, Yuseong-gu, Daejeon 34055, Republic of Korea}
\affiliation[b]{KASI Campus, University of Science and Technology, 217 Gajeong-ro, Yuseong-gu, Daejeon 34113, Republic of Korea}
\affiliation[c]{Department of Physics, TUM School of Natural Sciences, Technical University of Munich, James-Franck-Straße 1, 85748 Garching, Germany}
\affiliation[d]{Max-Planck-Institut fur Astrophysik, Karl-Schwarzschild-Str. 1, 85748 Garching, Germany}
\affiliation[e]{Energetic Cosmos Laboratory, Nazarbayev University, Astana 010000, Kazakhstan}
\affiliation[f]{Berkeley Center for Cosmological Physics, University of California, Berkeley, CA 94720, USA}
\affiliation[g]{Lawrence Berkeley National Laboratory, Berkeley, CA 94720, USA}
\affiliation[h]{Space Telescope Science Institute, 3700 San Martin Drive, Baltimore, MD 21218, USA}

\emailAdd{swpark@kasi.re.kr}
\emailAdd{shafieloo@kasi.re.kr}

\abstract{Internal kinematics of galaxies, traced through the stellar 
rotation curve or two dimensional velocity map, carry 
important information on galactic structure and dark 
matter. With upcoming surveys, the velocity map may play 
a key role in the development of kinematic lensing as an 
astrophysical probe. We improve techniques for extracting 
velocity information from integral field spectroscopy at 
low signal-to-noise ($S/N$), without a template, and demonstrate substantial 
advantages over the standard Penalized PiXel-Fitting method (pPXF) approach. Robust 
rotation curves can be derived down to $S/N\approx 2$ using our method. }


\begin{document}
\maketitle
\flushbottom

\section{Introduction}
\label{sec:intro}

The importance of the internal kinematics of galaxies in the fields of cosmology and astrophysics cannot be undermined. Galaxy dynamics are pivotal in understanding galactic structures, comprehending galaxy dynamics within clusters, and offering vital insights into unraveling the enigmas of dark matter. Dark matter, an essential component of the Universe, holds indispensable clues to solving numerous cosmological questions. However, its elusive nature, characterized by a lack of interaction with baryonic matter, poses significant challenges to its comprehension.

The existence of an anomalous gravitational field in and around galaxies was first suggested by Zwicky through observations of velocity dispersion for galaxies in the Coma cluster~\cite{1933AcHPh...6..110Z}. This was the first discovery of dark matter. Subsequently, discrepancies in the galaxy rotation curve compared with the optical stellar disk were reported, indicating the presence of an unknown source of gravitational field at the outskirts of galaxies~\cite{1939LicOB..19...41B}. Further studies of spiral galaxies confirmed the existence of dark matter at their outskirts~\cite{1970ApJ...159..379R, 1973A&A....26..483R, 1980ApJ...238..471R}.  As the galaxy rotation curve provides evidence for the presence of dark matter, the accurate estimation of these curves has become increasingly important in contemporary astrophysics.

There have been numerous significant discoveries in galaxy dynamics, including the Fundamental plane~\cite{1973AJ.....78..583G, Djorgovski:1987vx} and Fabor-Jackson relation~\cite{Faber:1976sn} for elliptical galaxies, and the Tully-Fisher relation~\cite{1977A&A....54..661T} for spiral galaxies, and so on. The measurements of line-of-sight (LOS) galaxy dynamics are important for understanding the dark matter distribution within galaxies.~\cite{1978ApJ...225L.107R, 1981AJ.....86.1825B, 1985ApJ...295..305V, 1996MNRAS.281...27P, 2011MNRAS.415..545T}. 

For observations, the Integral Field Unit Spectroscopy (IFU) has become increasingly important. IFUs offer spectral information across a spatially extended field of view, facilitating extended spectral observations of astronomical objects such as nebulae and galaxies. Several large-scale surveys have been conducted, including CALIFA (\href{https://califa.caha.es/}{https://califa.caha.es/})~\cite{2012A&A...538A...8S, 2016A&A...594A..36S}, SAMI (\href{https://sami-survey.org/}{https://sami-survey.org/})~\cite{2012MNRAS.421..872C,2021MNRAS.505..991C}, MaNGA (\href{https://www.sdss4.org/surveys/manga/}{https://www.sdss4.org/surveys/manga/})~\cite{2015ApJ...798....7B, 2015AJ....149...77D}, and KMOS (\href{https://www.mpe.mpg.de/ir/KMOS3D}{https://www.mpe.mpg.de/ir/KMOS3D/})~\cite{2015ApJ...799..209W, 2016MNRAS.457.1888S}, have compiled IFU databases of thousands of galaxies.

In traditional methods, determining galaxy kinematics involves matching spectra through spectral template fitting~\cite{Cappellari:2003hi, Fernandes2005, Ocvirk2006, Walcher:2006hd, Koleva:2009kt, 2017MNRAS.466..798C}. However, these methods heavily rely on templates, making it challenging to accurately estimate LOS velocities in low $S/N$ galaxies or on the outskirts of galaxies. For instance, the Penalized PiXel-Fitting Method (pPXF)~\cite{Cappellari:2003hi, 2017MNRAS.466..798C, 2023MNRAS.526.3273C}, which is a technique used in Spectral Energy Distribution (SED) fitting. It employs maximum penalized likelihood to determine the LOS velocity distribution. Nevertheless, this method also struggles with high uncertainty when dealing with low $S/N$ data.

Recently, \cite{Bag:2021xym} introduced a model independent approach for measuring LOS velocities using cross-correlation with iterative smoothing to calculate galaxy rotation curves. They showed the effectiveness of this method for high $S/N$ galaxy, while also suggesting its potential applicability for low $S/N$ galaxies. Following this, \cite{Denissenya:2023tzw} used a model independent method across a 2D array of spaxels (spectral pixels: datacube with angular position plus spectrum) from an integral field spectrograph to determine galaxy rotation curves. They improved the robustness of the method, enabling results for lower $S/N$ data, and extended this approach from 1D to 2D, successfully reconstructing a 2D galaxy stellar velocity map.

In this paper, we present an enhanced template-free technique designed to calculate the velocity differences between a pair of spaxels. This method utilizes iterative smoothing and cross-correlation techniques applied to IFU data spectra, as discussed in~\cite{Bag:2021xym}. Specifically, we aim to address the challenges posed by high noise data in reconstructing galaxy rotation curves, focusing on extremely low $S/N$ MaNGA galaxies. By estimating Doppler shifts through the cross-correlation of spectra, along with both smoothing and the selection of the particular parts of spectra, we successfully calculate the rotation curves for low $S/N$ galaxies. Our method offers several advantages over standard methods as it does not require templates and is not dependent on specific individual features. Furthermore, our approach demonstrates outstanding performance, particularly for low $S/N$ galaxies, compared to other methods such as pPXF (\href{https://pypi.org/project/ppxf/}{https://pypi.org/project/ppxf/}) and Marvin (\href{https://pypi.org/project/sdss-marvin/}{https://pypi.org/project/sdss-marvin/}), the tool which is used to analyze and visualize the MaNGA data~\cite{2019AJ....158...74C}. 
Additionally, we validate the consistency between our results and other results obtained for the high $S/N$ galaxy. In summary, our method exhibits robust performance regardless of $S/N$ levels.

This paper is organized as follows. In Section~\ref{sec:Methodology}, we explain the novel approach used in this paper, comprising iterative smoothing and cross-correlation, and apply these techniques to analyze low $S/N$ MaNGA galaxies. Following that, we present results and analysis based on galaxy rotation curves for each galaxy in Section~\ref{sec:results}. In Section~\ref{sec:ppxf}, we compare our rotation curves with the traditional pPXF results. Finally, we conclude and discuss future works in Section~\ref{sec:conclusion}.

\begin{table}
\caption{\label{tab:table1} The details of 10 low $S/N$ MaNGA galaxies used in this paper are presented, with the galaxies sorted by their median $S/N$. For galaxy $\#6$, no redshift information is available in the MaNGA database. Galaxy $\#0$ is included as a ``Good'' quality galaxy for comparison purposes.} 
\,\\ 
\centering
\begin{tabular}{|c|c|c|c|c|c|}
\hline
No. & Name & $z$ &median $S/N$ \\
\hline

$\#1$ & 12495-12704 & 0.0450 & 5.55  \\
$\#2$ & 11950-12703 & 0.0966 & 4.46  \\
$\#3$ & 12495-12703 & 0.0689 & 3.87  \\
$\#4$ & 11950-12701 & 0.0413 & 3.53  \\
$\#5$ & 12490-12703 & 0.0386 & 3.00  \\
$\#6$ & 11950-12705 & - & 2.55  \\
$\#7$ & 12491-12702 & 0.0390 & 2.44  \\
$\#8$ & 12490-9102 & 0.0392 & 1.97 \\
$\#9$ & 12495-3702 & 0.0430 & 1.94  \\
$\#10$ & 12495-12701 & 0.1219 & 1.64 \\
$\#0$ & 7991-12701 & 0.0289 & 27.8  \\
\hline
\end{tabular}
    
\end{table}

\section{Methodology}
\label{sec:Methodology}

The aim is to map out the rotation curves of galaxies, i.e., the differential velocity as they rotate around their center. By observing the flux of the galaxy through a spectrograph, we can gather information about each spaxel in the 2D array. When comparing one spaxel to another, we can detect LOS wavelength shifts due to the Doppler effect caused by the rotation of the galaxy. For instance, the spaxel moving away from the observer will have a redshift compared to the center, while the opposite side of the spaxel will have a blueshift. Consequently, we can determine the velocity difference $\Delta V$ relative to the central spaxel.

In this paper, we analyze 10 low $S/N$ MaNGA galaxies using iterative smoothing and cross-correlation methods to estimate the galaxy rotation curve as suggested in~\cite{Bag:2021xym}. Table~\ref{tab:table1} presents the details of the selected 10 low $S/N$ galaxies used in this paper, including their number, name, redshift, and median $S/N$. The galaxies are sorted by their median $S/N$, and henceforth we will refer to each galaxy by its number. For example, $\#2$ 11950-12703 is referred to simply as $\#2$. \cite{Bag:2021xym} presents a novel approach employing iterative smoothing followed by spaxel cross-correlation, which was validated using simulated data and applied to a real galaxy, and further tested in \cite{Denissenya:2023tzw}. To estimate the rotation curve, we select an axis including the central spaxel in the 1D array, typically the galaxy major axis is selected. 

In the next subsections, we introduce the details of iterative smoothing, cross-correlation to calculating $\Delta V$, and some technical aspects for applying them to low $S/N$ galaxies.

\begin{figure}

\centering

\includegraphics[width = 0.8\linewidth]{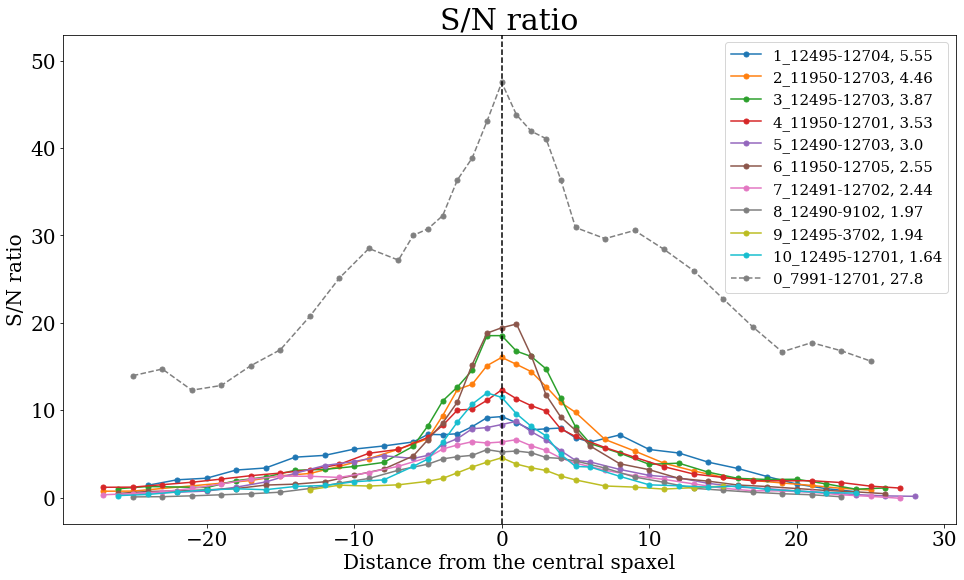}
\caption{\label{fig:fig_SN} $S/N$ ratio for fluxes in the spaxels is presented for 10 different low $S/N$ galaxies, as listed in Table~\ref{tab:table1}. 
Labels indicating the names sorted by the median $S/N$ for individual galaxies (last number in the label). Galaxy $\#0$ is included as a reference for a ``Good'' quality galaxy. All of our other galaxies have a low median $S/N\le5.55$.}  
  
\end{figure}

\subsection{Iterative smoothing for the galaxy spectra}
\label{sec:M1}

Traditional fitting methods like pPXF fit raw, noisy stellar population spectra using model stellar templates. The raw spectra include both stellar absorption lines and gas emission lines. For this analysis, we use the complete spectra, which contain both stellar and gas components, enabling us to obtain the rotation curves from the combined stellar and gas velocities. While in this paper we used the complete spectra of the galaxies in our analysis, it is possible to study the effect of different components separately. In Appendix~\ref{sec:appendix_b}, we demonstrate how our method performs when we mask the gas emission lines to test the consistency of our results. While the separation of the spectral components is beyond the scope of this paper, it would be an important and interesting exercise, especially in cases where the stellar and gas spectral components of galaxies exhibit different behaviors. We will study such galaxies in our future works.

Our template-free method uses both raw (unsmoothed) and smoothed spectra; raw spectra shapes are often unsuitable for direct cross-correlation due to observational noise, which can sometimes obscure the true wavelength shift from the Doppler effect. Thus, we have adopted the iterative smoothing method using a Gaussian kernel, as described in~\cite{Shafieloo:2005nd, Shafieloo:2007cs, Shafieloo:2009hi, Aghamousa:2014uya}. This smoothing renders our method less sensitive to noise and eliminates the reliance on strong spectral features.

The smoothed spectrum obtained in the $n$-th iteration step from the previous $(n-1)$-th step is given by
\begin{equation}
F^{s}_{n}(\lambda)=F^{s}_{n-1}(\lambda)+N(\lambda) \sum_{i} \dfrac{F_{\textrm{obs}}\left(\lambda_{i}\right)-F^{s}_{n-1}\left(\lambda_{i}\right)}{\sigma_{\textrm{obs}}^{2}\left(\lambda_{i}\right)} \times \exp \left[-\dfrac{\left(\lambda_{i}-\lambda\right)^{2}}{2 \Delta^{2}}\right],
\label{eq:iter_1}
\end{equation}

where
\begin{equation}
N(\lambda)^{-1}=\sum_{i} \exp \left[-\dfrac{\left(\lambda_{i}-\lambda\right)^{2}}{2 \Delta^{2}}\right] \dfrac{1}{\sigma_{\textrm{obs}}^{2}\left(\lambda_{i}\right)}\ .
\label{eq:iter_2}
\end{equation}

Here, $F_{\textrm{obs}}(\lambda_{i})$ and $\sigma_{\textrm{obs}}(\lambda_{i})$ represent the $i$-th data point and its uncertainty. $F^{s}_{n-1}(\lambda)$ is the smoothed spectrum calculated from the previous iteration, starting with a first guess model $F^{s}_{0}(\lambda)$. $N(\lambda)$ is a normalization factor, and $\Delta$ is the width of smoothing. 

In this cycle, the first iteration begins with $F^{s}_{0}(\lambda)$ = constant. The result of the first smoothing is $F^{s}_{1}(\lambda)$, the next iteration calculates $F^{s}_{2}(\lambda)$ and so on. With a sufficient number of iterations $N_{\textrm{it}}$ the results tend to be independent of the initial guess model, and moreover the smoothed fit generally captures the essential features in the actual data at the applied smoothing scale.

The smoothing scale $\Delta$ plays an important role in handling the noise for low $S/N$ galaxies. In Section~\ref{sec:M3}, we discuss the impact of different choices of $\Delta$ values on the final results.

\subsection{Calculating $\Delta V$ using cross-correlation}
\label{sec:M2}

 To obtain the LOS velocity difference $\Delta V$ between a pair of spaxels, we cross-correlate the two spectra at multiple wavelength shifts. We expect an excess power in the correlation coefficient when the wavelength shift matches the Doppler shift. To preserve the maximum information in the data, while handling the noise, we employ one raw and one smoothed spectrum. The weighted cross-correlation coefficient between the spectra of two spaxels $A$ and $B$ is

\begin{equation}
r_{A^s B}(\Delta V) \equiv F_A^s(\lambda+\Delta \lambda) \otimes F_B(\lambda)= \\
\\
    \frac{\sum_i w_i\, \Delta F^s_A(\lambda_i+\Delta \lambda_i)\,\Delta F_B(\lambda_i)}
{\sqrt{\sum_i w_i \left[\Delta F^s_A(\lambda_i+\Delta \lambda_i)\right]^2}\sqrt{\sum_i w_i \left[\Delta F_B(\lambda_i)\right]^2}}\ , 
\label{eq:cc}
\end{equation}
where $\Delta \lambda_i = \lambda_i\,(\Delta V/c)$, and $\Delta F = F - \langle F\rangle_w$. We use inverse variance weights $w_i$ = 1/$\sigma_{Bi}^2$, where $\sigma_{Bi}$ is the uncertainty of unsmoothed spectrum $B$, and the index $i$ runs over data points in wavelength. The maximum of the correlation function determines $\Delta V$. 

The shifted smoothed spectrum $F_A^s(\lambda_i + \Delta \lambda_i)$ is derived from the original spectrum $F_A(\lambda_i)$ as described in the previous subsection. Again note only one spectrum is smoothed using Eq.~(\ref{eq:iter_1}), while the other is left unaltered. As a form of cross-validation, we then interchange which of the pair is smoothed. That is, at first we cross-correlate the shifted smoothed spectrum $F_A^s$ with the unsmoothed spectrum $F_B$, and get $\Delta V_{AB}$. Then we interchange their roles and do the same calculations to get $\Delta V_{BA}$, so we obtain the LOS velocity difference between $F_B^s$ and $F_A$. The correlation functions $r_{A^s B}$ and $r_{B^s A}$ should ideally have their maximum at the same $\Delta V$ but with opposite signs, i.e. 
\begin{equation}
\Delta V_{AB} \approx -\Delta V_{BA}\ .
\label{eq:cc2}
\end{equation}


\subsection{$\Delta V$ estimation for low $S/N$ galaxies}
\label{sec:M3}

\begin{figure}

\centering

\subfloat[15 bins]{\includegraphics[width = 0.48\linewidth]{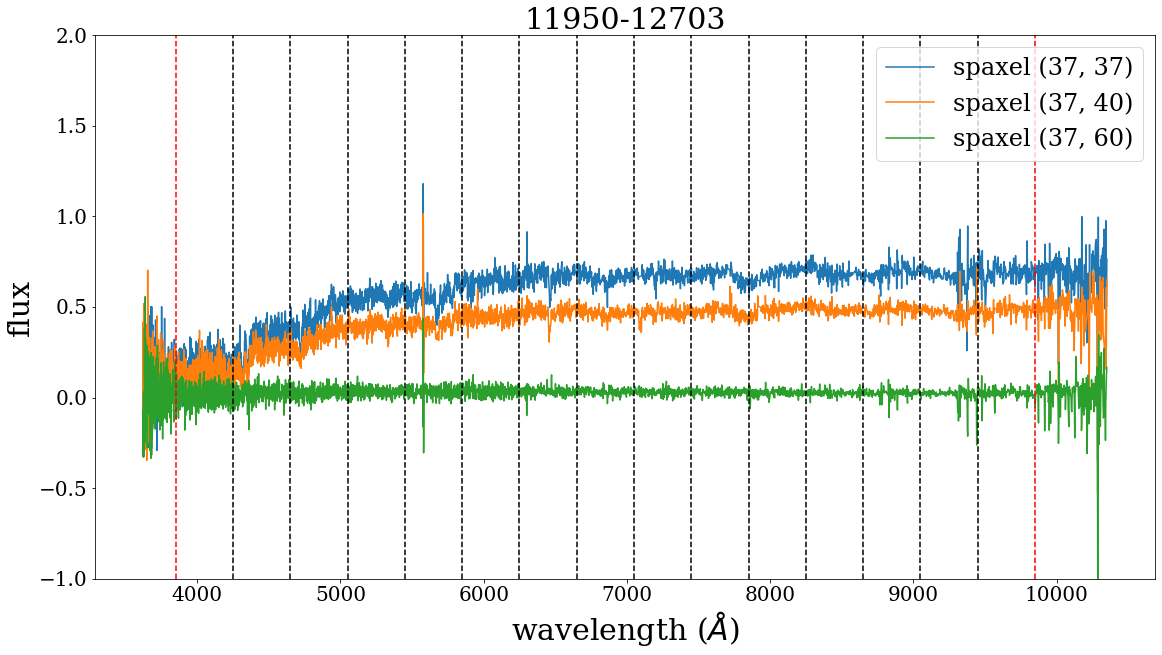}}
    \label{fig:15bins_spectra}
\subfloat[30 bins]{\includegraphics[width = 0.48\linewidth]{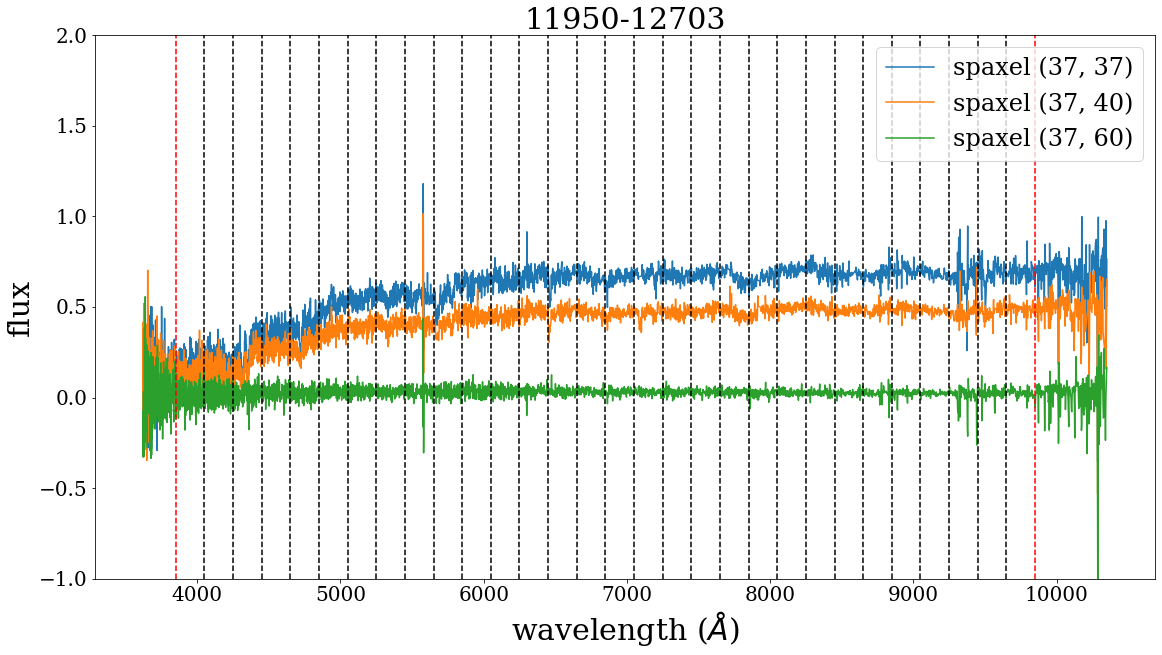}}
    \label{fig:30bins_spectra}

\caption{\label{fig:fig_spectra} The spectra of MaNGA galaxy $\#2$ are shown for three spaxels, labeled by the 2D position in the IFU grid, with (37, 37) representing its central spaxel.  Close to the central spaxel, the spectra show similar shapes, but these can differ for spaxels further away. The entire range is divided into 15 bins (\textit{left}) and 30 bins (\textit{right}), with contaminated regions $\lambda$ $<$ 3850 \AA{} and $\lambda$ $>$ 9850 \AA{} discarded.}  
  
\end{figure}

\begin{figure}

\centering

\subfloat[37 $\otimes$ 40]{\includegraphics[width = 0.48\linewidth]{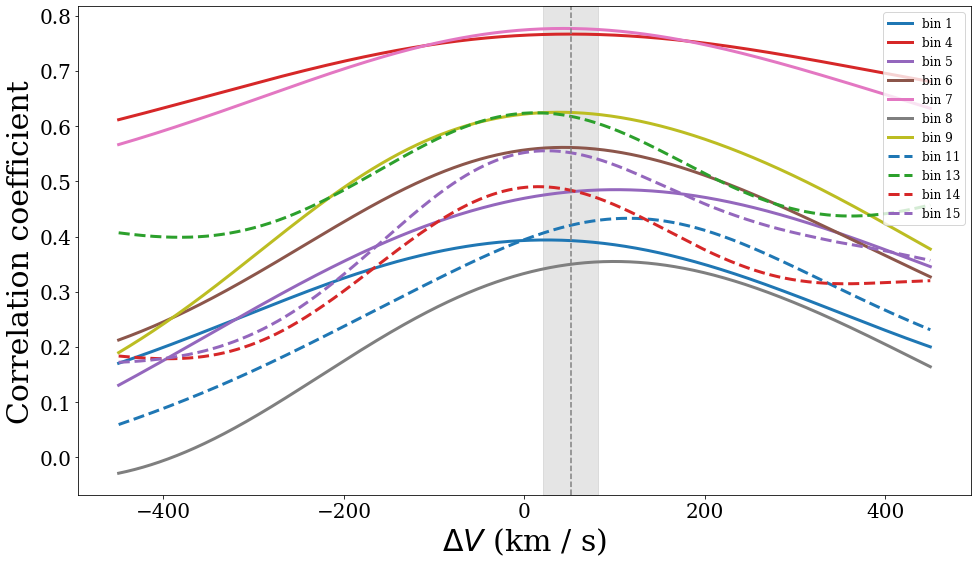}}
    \label{fig:a_2}
\subfloat[40 $\otimes$ 37]{\includegraphics[width = 0.48\linewidth]{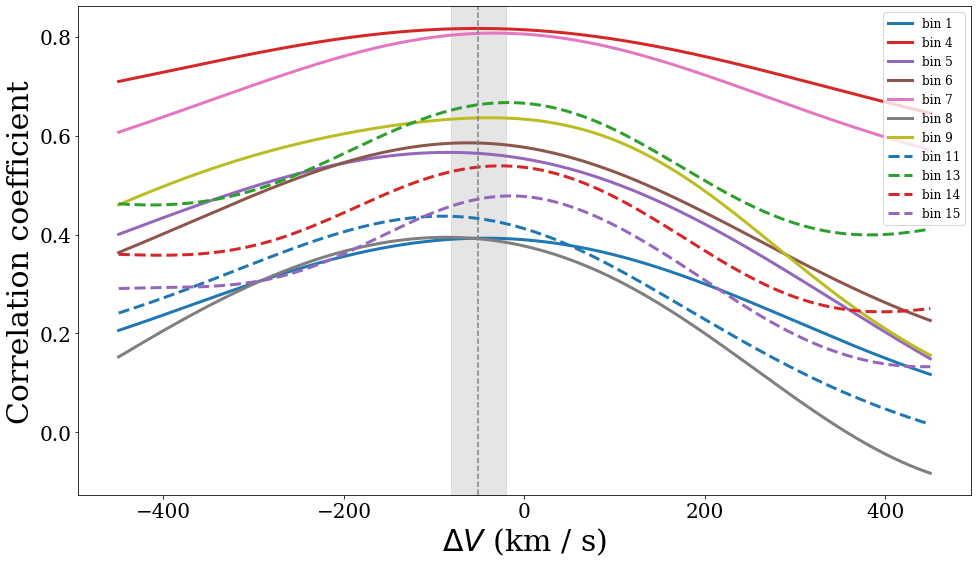}}
    \label{fig:b_2}

\subfloat[37 $\otimes$ 56]{\includegraphics[width = 0.48\linewidth]{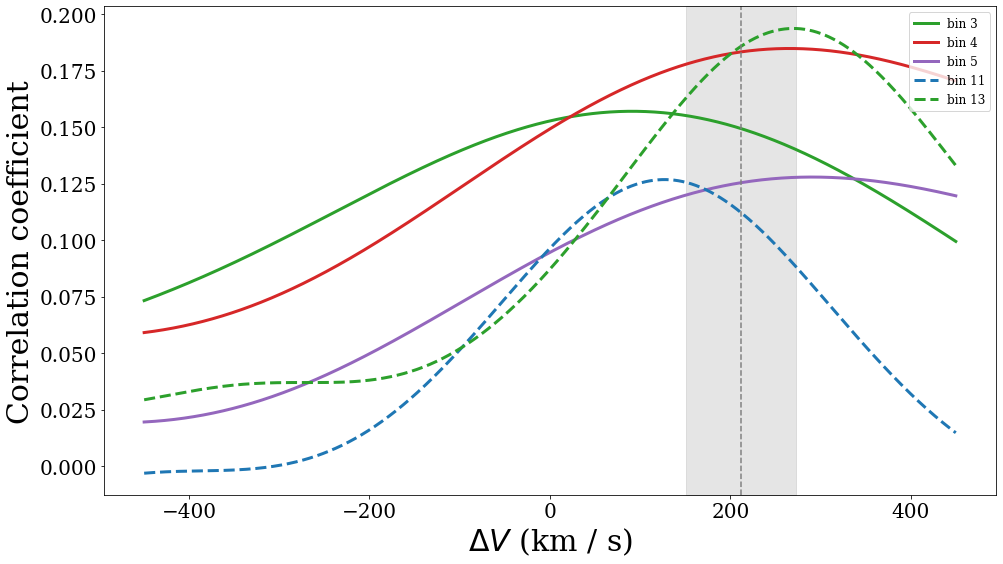}}
    \label{fig:c_2}
\subfloat[56 $\otimes$ 37]{\includegraphics[width = 0.48\linewidth]{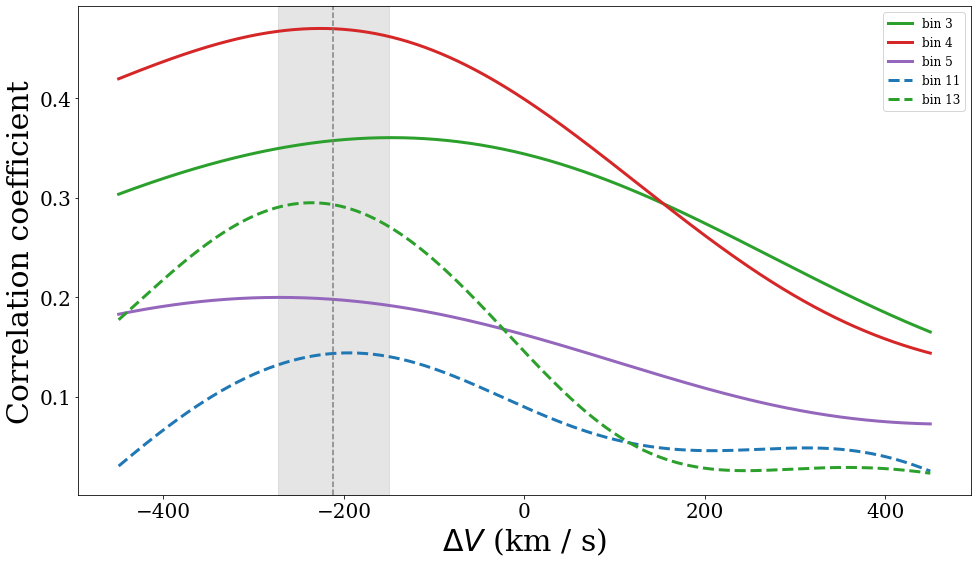}}
    \label{fig:d_2}

\caption{\label{fig:fig_correlation} The correlation coefficient of galaxy $\#2$ is shown as a function of trial velocity difference $\Delta V$ with a smoothing scale of $\Delta = 6.0$ \AA. 
Two example spaxel pairs and their inverses (smoothed on the other member of the pair) are shown; spaxel $(37,N)$ is simply referred to as spaxel $N$. The consistency criteria check that the correlation coefficient for a given pair and its inverse agree (with a minus sign); 11 wavelength bins pass for the inner spaxel (40) while only 5 pass for the outer (56) where the $S/N\approx1$. The gray region is the estimated velocity and its uncertainty from the cross-validation check. 
}  
  
\end{figure}

Our major goal is to extend the cross-correlation and iterative smoothing techniques for calculating galaxy rotation curves to lower $S/N$ galaxies. In \cite{Bag:2021xym} the method demonstrated its effectiveness by analyzing a MaNGA galaxy ($\#0$ in Table \ref{tab:table1}). In the Marvin database, galaxy $\#0$ is classified as ``Good'' quality data according to MaNGA criteria and exhibits a high $S/N$, with $S/N$ values of spaxels defined following \cite{2008ASPC..394..505S}. Shown in Figure~\ref{fig:fig_SN}, the $S/N$ for this galaxy has a maximum of 47.5, and a median of $\sim27$. In \cite{Denissenya:2023tzw}, the enhanced technique was again applied to this galaxy and applied to simulations of lower $S/N$ galaxies. 

Here, to further evaluate our model independent method's performance for galaxies with low $S/N$ ratios, we select 10 additional MaNGA galaxies, all of which are labeled with a``Warning'' label in the MaNGA criteria. All selected MaNGA galaxies have a median $S/N$ below 5.6, with spaxels away from the center approaching $S/N\approx1$. 

Therefore we adjust the technique with respect to spectral bins and velocity consistency criterion. 
In \cite{Bag:2021xym,Denissenya:2023tzw}, the entire spectra were divided into 4 wavelength bins for galaxy $\#0$, and the smoothing scale is set to $\Delta = 1.5$ \AA{} for iterative smoothing. Additionally, they imposed the following consistency criteria to select the good bins for cross-correlation: 
\begin{equation}
|\Delta V_{AB} + \Delta V_{BA}| \leq 0.05|\Delta V_{AB} - \Delta V_{BA}|, \ \textrm{or} \leq 10\ \textrm{km/s}.
\label{eq:cri1}
\end{equation}

The velocity estimation is calculated with respect to the central spaxel, and whichever is the looser criterion of the two is adopted. 

For low $S/N$ galaxies these criteria are too restrictive, making it challenging to find bins that fit into the criteria. Moreover, determining the maximum cross-correlation values becomes difficult if the pair of spectra differ too much in shape, due to noise plus stellar population shifts in different galaxy regions. Therefore, we divide the spectra into 15 and 30 wavelength bins for more effective cross-correlation results. Figure~\ref{fig:fig_spectra} illustrates divided spectra range into 15 and 30 bins respectively. In addition, we relax the consistency criteria for low $S/N$ galaxies to 
\begin{equation}
|\Delta V_{AB} + \Delta V_{BA}| \leq 0.3|\Delta V_{AB} - \Delta V_{BA}|, \ \textrm{or} \leq 20\ \textrm{km/s}.
\label{eq:cri2}
\end{equation} 

Although we divide numerous bins and relaxed selection criteria from Eq.~\ref{eq:cri2} to calculate cross-correlation, estimating galaxy rotation curves remains challenging. For instance, it is still quite hard to estimate the velocity of the outskirts of the galaxy because of too low $S/N$. 

Figure~\ref{fig:fig_correlation} illustrates the cross-correlation coefficient for galaxy $\#2$, showing only the bins that passed the consistency criteria of Eq.~\ref{eq:cri2}. To achieve cross-correlation with good bins, we initially examine the trend of the rotation curve around the central spaxel where the $S/N$ is relatively high, to determine if the rotation direction is clockwise or counterclockwise. 

We should note here that there are galaxies with counter-rotating inner and outer regions and for such galaxies, results can become confusing and unreliable if one forces a general form of templates for the galaxy rotation curves. In our analysis we found no evidence of such cases in the galaxies we studied. The spaxels in the outer regions show consistency with the rotation direction inferred from the inner regions, supporting a uniform rotation throughout the galaxies. If a galaxy exhibited counter-rotating discs, our analysis would detect discrepancies in the derived velocities ($\Delta V$) between the inner and outer spaxels. We should emphasize that we have been using five anchor points in the inner region, and such discrepancies in the derived $\Delta V$s could be more evident if we had such a special galaxy in our sample. While this would be an interesting case to investigate, it falls beyond the scope of our current study. The details for using five anchor spaxels will be discussed later in this section.\\

Subsequently, based on this determination, we selectively choose the bins exhibiting the correct direction, as shown in Figure~\ref{fig:fig_correlation}. For instance, if $\Delta V$ is positive for spaxel (40), we exclusively consider positive $\Delta V$ values for larger spaxels, such as for spaxel (56). After cross-validation with the criteria based on Eq.~\ref{eq:cri2}, the average value between $\Delta V_{AB}$ and $-\Delta V_{BA}$ will represent the $\Delta V_{AB}$ of each bin. The mean and standard deviation of all the passed bins will estimate final $\Delta V_{AB}$ and uncertainty respectively, as shown in the gray region of Figure~\ref{fig:fig_correlation}.

We implement several crosschecks to establish the velocity through the cross-correlation. These include consistency criteria upon switching which spectra of the pair are smoothed, multiple wavelength regions for independent analysis, and consistency between the 15 bins and 30 bins analysis. We further require that more than 25\% of wavelength bins pass to ensure reliable results. In other words, when there are 15 bins and 30 bins, only cases where at least 4 bins and 8 bins passed are selected respectively. For example, in Figure~\ref{fig:fig_correlation}, spaxels (40) and (56) are considered reliable results because 11 and 5 out of 15 bins passed, respectively. In situations where fewer bins pass our criteria, we utilize interpolation. For instance, if spaxel $(N)$ only passes 3 bins, which is insufficient to meet our criteria, we omit the calculation of $\Delta V$ for spaxel $(N)$ and instead interpolate this point between spaxels $(N-1)$ and $(N+1)$.

As expected, $\Delta V$ (relative to the center spaxel 37) for outer spaxels with lower $S/N$ becomes more challenging to measure. We see that 11 out of 15 bins pass the criteria for the inner spaxel (40), whereas only 5 passed for the outer spaxel (56). 

While calculating $\Delta V$ for all pairs of spaxels relative to each spaxel might seem ideal, it introduces a challenge due to the high correlation between them. This correlation implies that calculated $\Delta V_{AB}$ can significantly influence the velocity estimation not only for spaxel $V_A$ but also for spaxel $V_B$. Hence, spaxels in regions with lower $S/N$ might yield inaccurate results or introduce bias, affecting not only their own velocity estimations but also those of the higher $S/N$ spaxels. To mitigate this issue, we designate anchor spaxels around the central spaxel, usually in the highest $S/N$ region, and consider the velocities of all spaxels with respect to only each of the anchor spaxels in our final calculation. This way we ensure that the low $S/N$ outer spaxels have minimal influence on the velocity estimation of inner spaxels, thereby reducing bias in the final calculation. 

We set five anchor spaxels including the center spaxel, with $\pm$ 2 intervals around the central spaxel for each galaxy to calculate the galaxy rotation curve. For instance, if the center spaxel for galaxy $\#2$ is 37, then 33, 35, 37, 39, and 41 could be the anchor spaxels, resulting in a total of five spaxels being used. We cross-correlate all the spaxels with each of these anchors. In other words, we determine the LOS velocity difference between the spaxels $i$ and $j$, i.e., $\Delta V_{ij}$, where $j$ runs over all anchor spaxels, and $i$ runs over all spaxels. Subsequently, we employ the Hamiltonian Monte Carlo (HMC) method with the \texttt{pystan} package~\cite{JSSv076i01} to obtain the final galaxy rotation curves. $\Delta V$ and uncertainties are the mean and standard deviation of the HMC results respectively. 

One may notice that the spectra in the outer regions can visually differ significantly in shape from those of the anchor spaxels. If the spectra in the outer regions are completely different (being uncorrelated) from those of the central anchor spaxels, the cross-correlation should not yield high correlation coefficients. In cases where we do observe relatively high correlations, it indicates that some features in the spectra of the inner and outer regions are similar. It is important to note that at least a few bins in the spectra (comparing two spaxels) should show high correlation coefficients simultaneously so that we can accept the derived $\Delta V$ for the next step of the analysis. This cannot occur between spaxels without intrinsic similarities.

To illustrate the impact of spectral differences, we refer to Figure~\ref{fig:fig_spectra}, which shows the spectra from MaNGA galaxy \#2. The central spaxel (37, shown in blue) and an inner spaxel (40, shown in orange) are visibly different from the outer spaxel (60, shown in green), but there are still some intrinsic similarities, maybe not so visible by eye, that our method can capture to calculate the $\Delta V$ between these spaxels (following our binning approach) so that we can derive the rotation curve of this galaxy subsequently up to its outskirts.

Moreover, by designating five anchor spaxels, we minimize the risk of false positives, reducing the likelihood of features from different spaxels randomly matching each other. For galaxies with significantly different stellar populations and gas properties at the center and outskirts, our approach might struggle to find any high correlation between the spaxels in these two regions and consequently fail to derive the rotation curve. However, in this paper, we have not encountered such a case, even though some of the galaxies have a very low $S/N$ in their spectra.

Furthermore, a short smoothing scale $\Delta$ = 1.5 \AA\  does not work well for low $S/N$ galaxies due to noise fluctuations. We tested a number of values for $\Delta$ and found that $\Delta=6.0$ \AA\ worked best. It will be discussed further in Appendix~\ref{sec:appendix}.


\begin{figure*}
\centering
    
\subfloat[Galaxy \#1]{\includegraphics[height=0.22\linewidth]{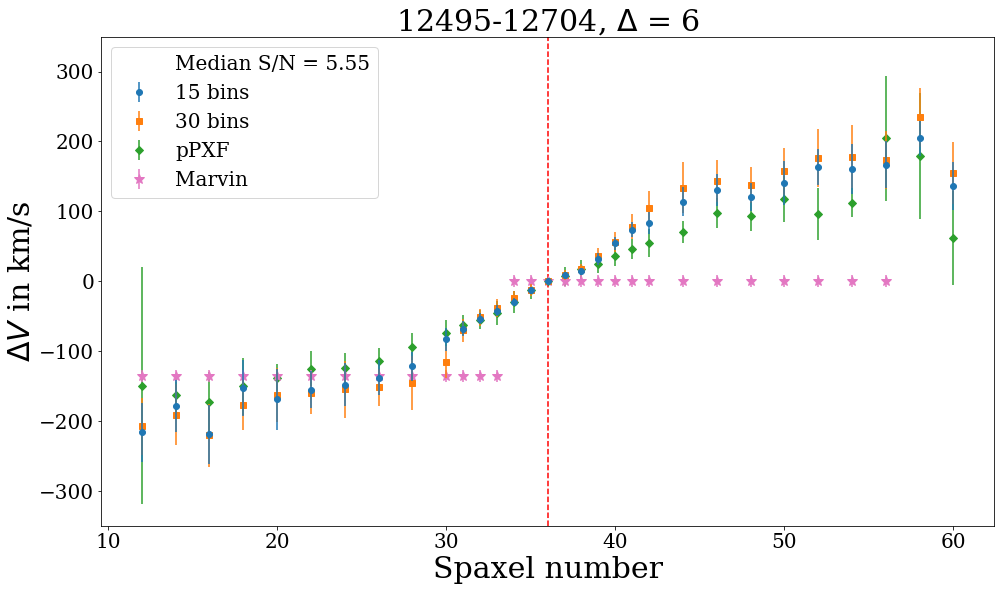}}
    \label{fig:a}
\hspace{1em}
\subfloat[Galaxy \#2]{\includegraphics[height=0.22\linewidth]{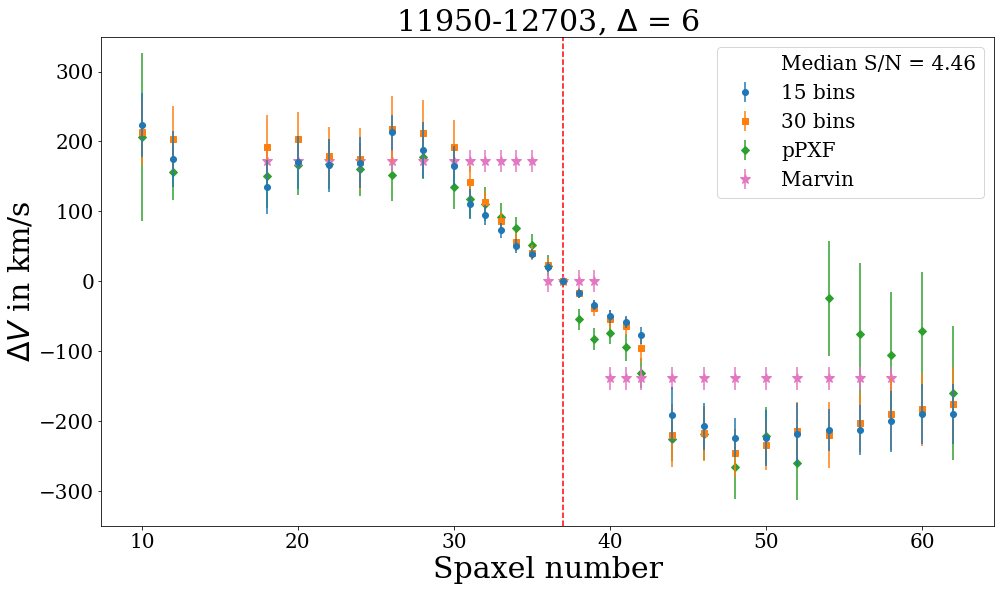}}
    \label{fig:b}

\subfloat[Galaxy \#3]{\includegraphics[height=0.22\linewidth]{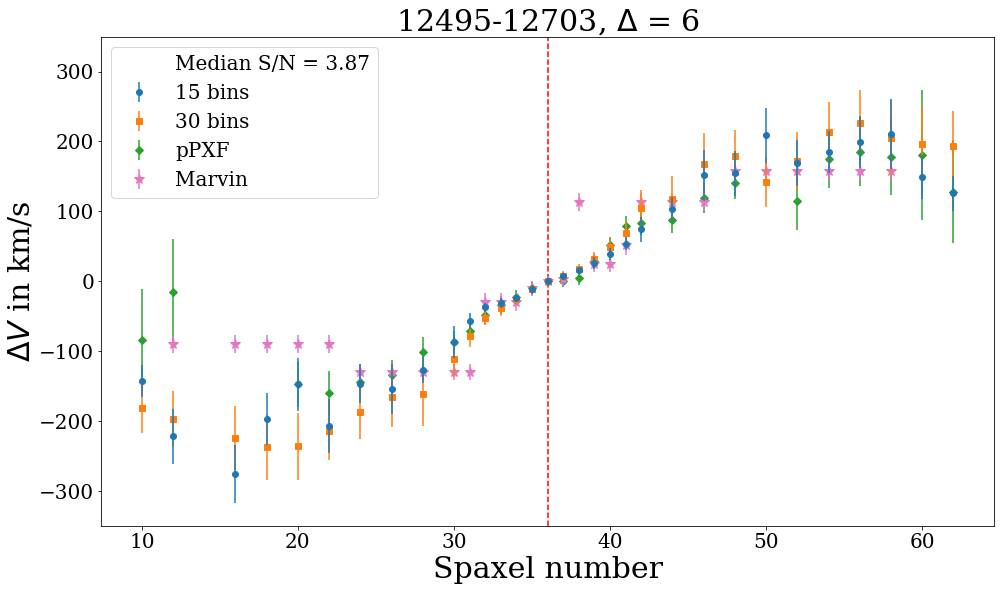}}
    \label{fig:c}
\hspace{1em}
\subfloat[Galaxy \#4]{\includegraphics[height=0.22\linewidth]{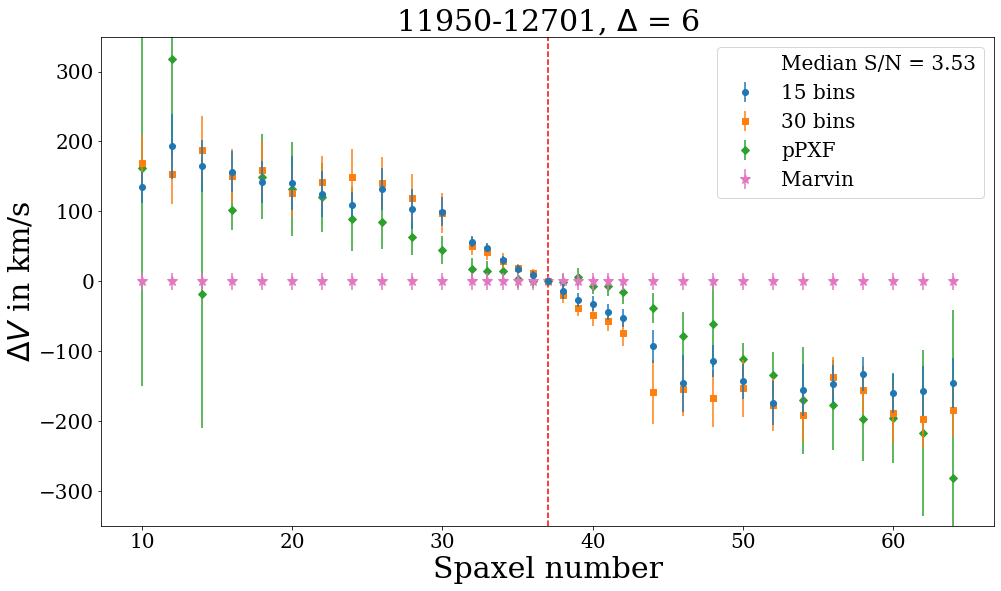}}
    \label{fig:d}

\subfloat[Galaxy \#5]{\includegraphics[height=0.22\linewidth]{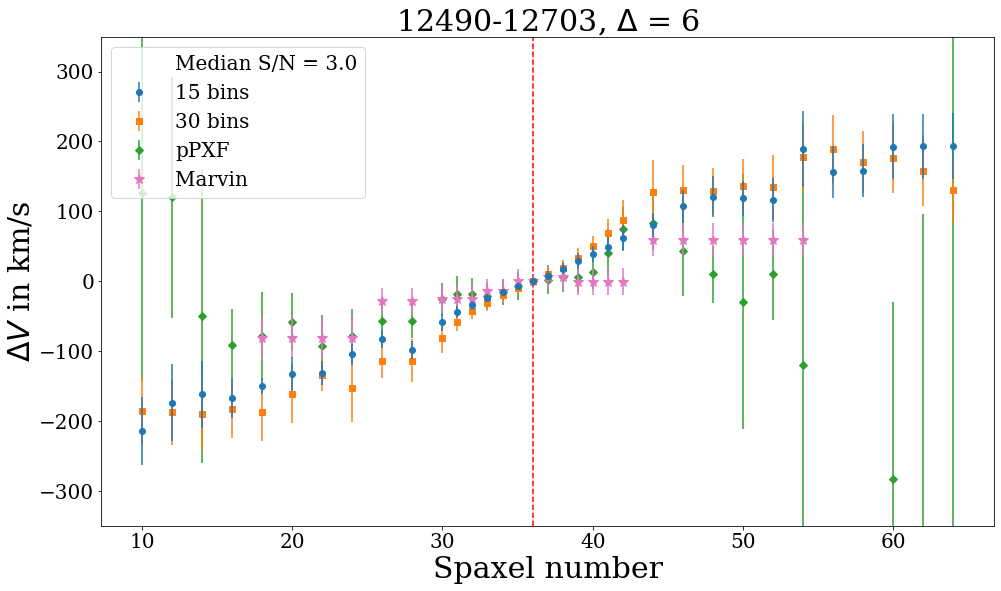}}
    \label{fig:e}
\hspace{1em}
\subfloat[Galaxy \#6]{\includegraphics[height=0.22\linewidth]{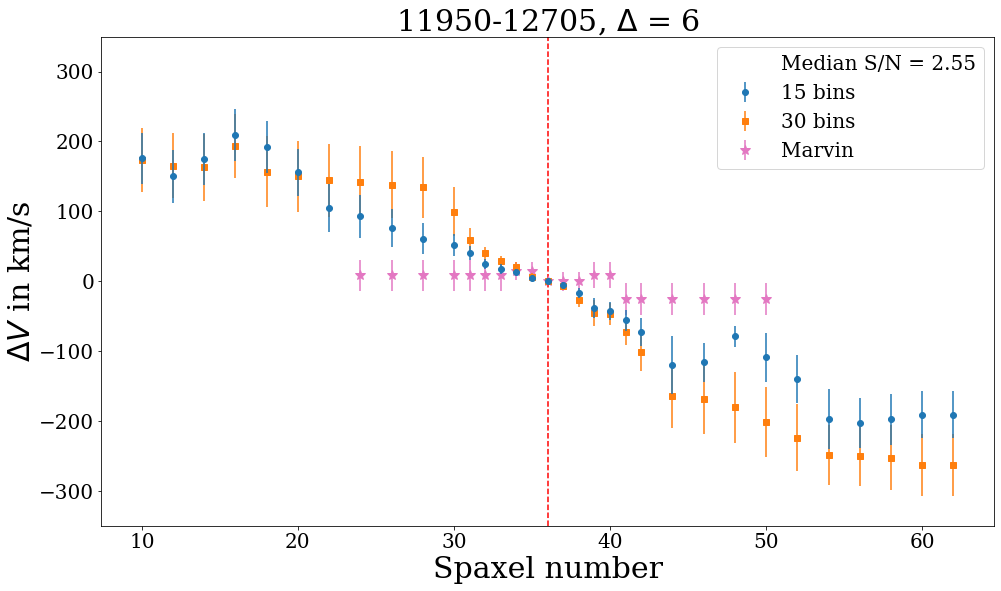}}
    \label{fig:f}

\subfloat[Galaxy \#7]{\includegraphics[height=0.22\linewidth]{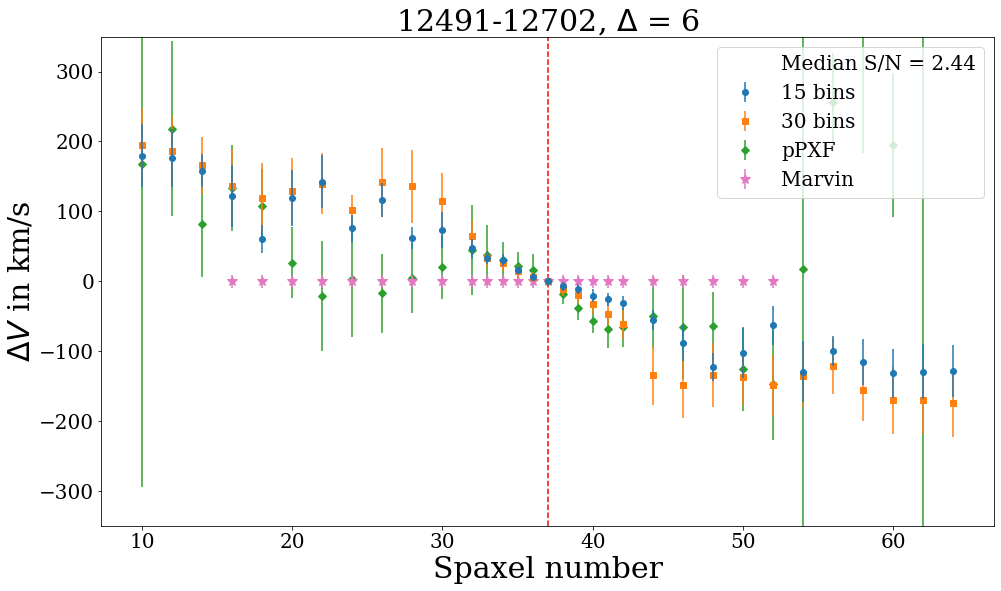}}
    \label{fig:g}
\hspace{1em}
\subfloat[Galaxy \#8]{\includegraphics[height=0.22\linewidth]{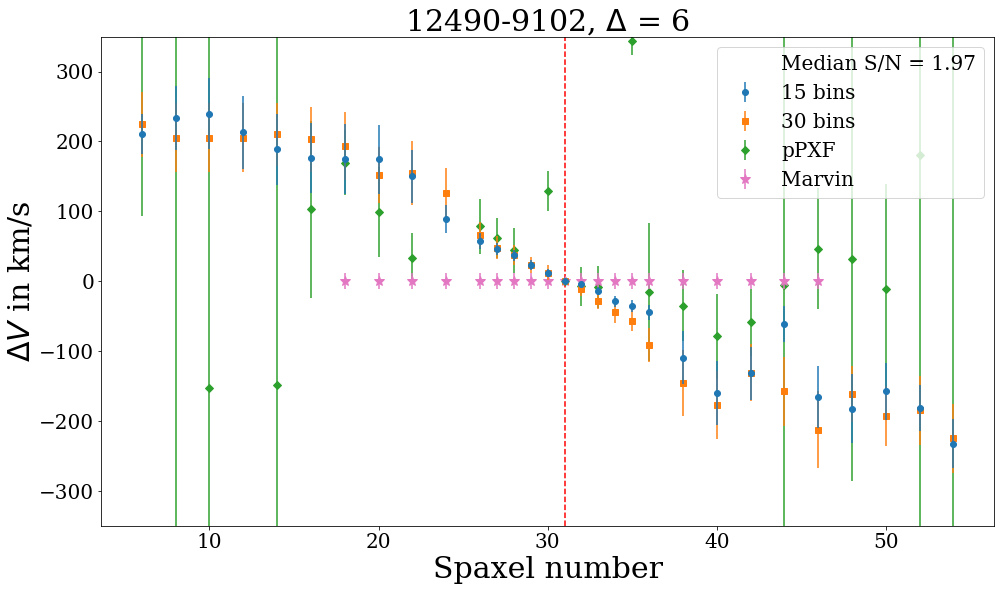}}
    \label{fig:h}

\subfloat[Galaxy \#9]{\includegraphics[height=0.22\linewidth]{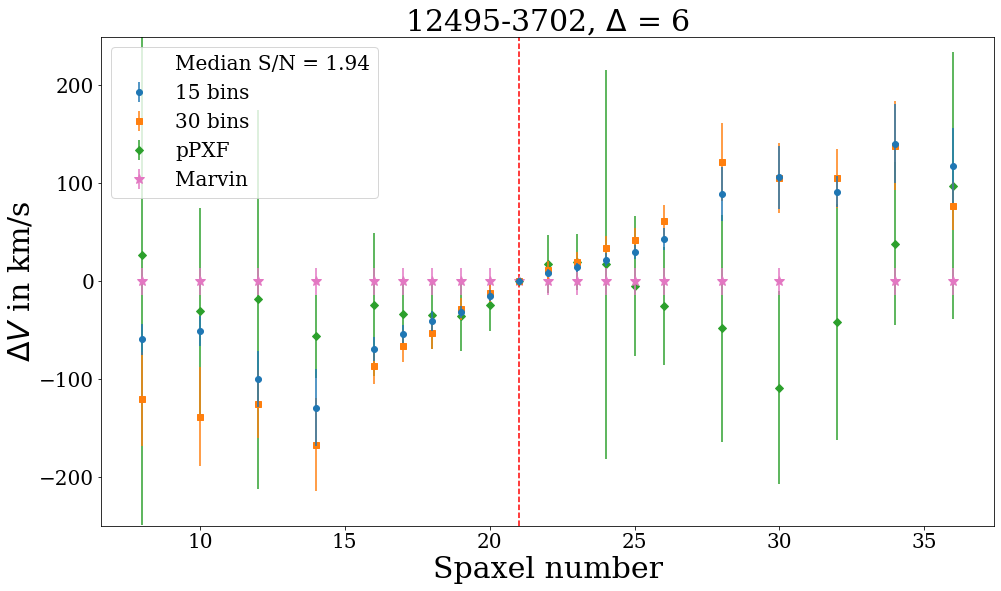}}
    \label{fig:i}
\hspace{1em}
\subfloat[Galaxy \#10]{\includegraphics[height=0.22\linewidth]{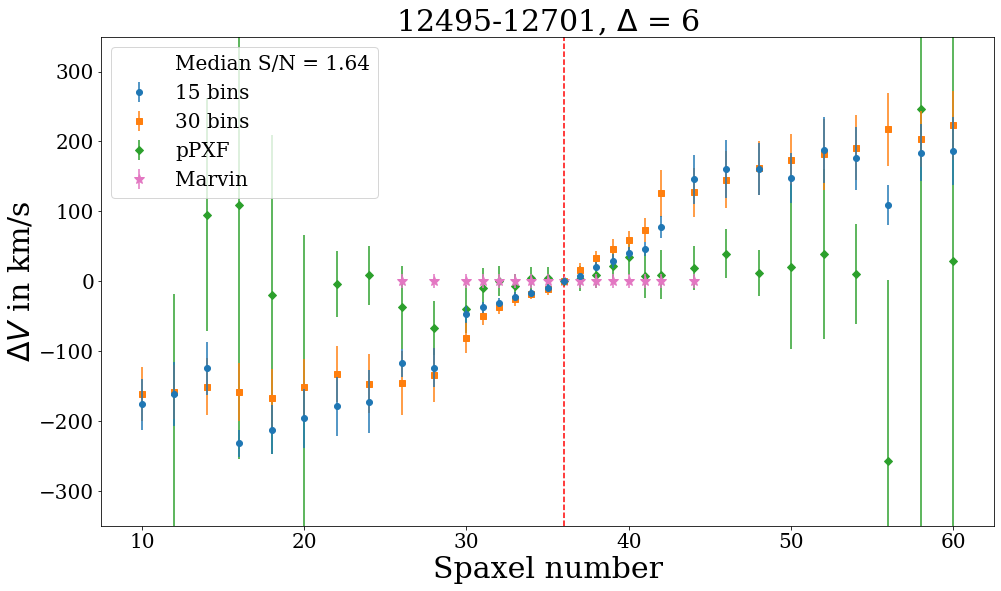}}
    \label{fig:j}

\caption{\label{fig:fig4} 
Galaxy rotation curves reconstructed for 10 low $S/N$ MaNGA galaxies, shown for our method with 15 (\textit{blue}) and 30 (\textit{orange}) wavelength bins. 
We also plot the stellar velocity estimation for the comparison of traditional pPXF (\textit{green}) and Marvin results (\textit{pink}). 
(Note for galaxy $\#6$, no redshift information is available in the MaNGA database, hence no pPXF result is shown.) 
In this low $S/N$ regime, the method presented here appears robust, while pPXF and Marvin begin to exhibit pathologies. 
}

\end{figure*}

\begin{figure*}
\centering
    
\subfloat[Galaxy \#1]{\includegraphics[height=0.22\linewidth]{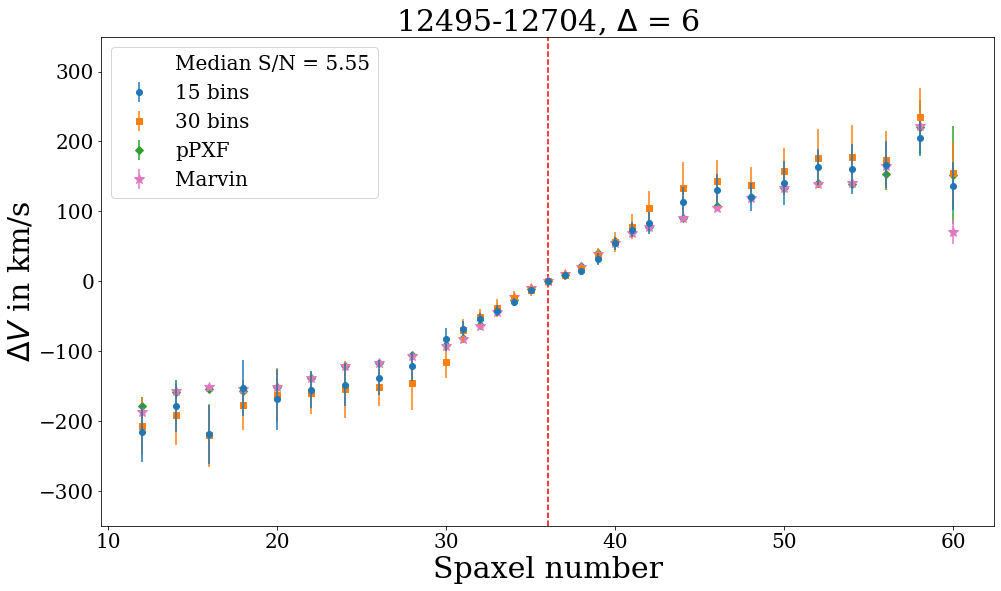}}
    \label{fig:a2}
\hspace{1em}
\subfloat[Galaxy \#2]{\includegraphics[height=0.22\linewidth]{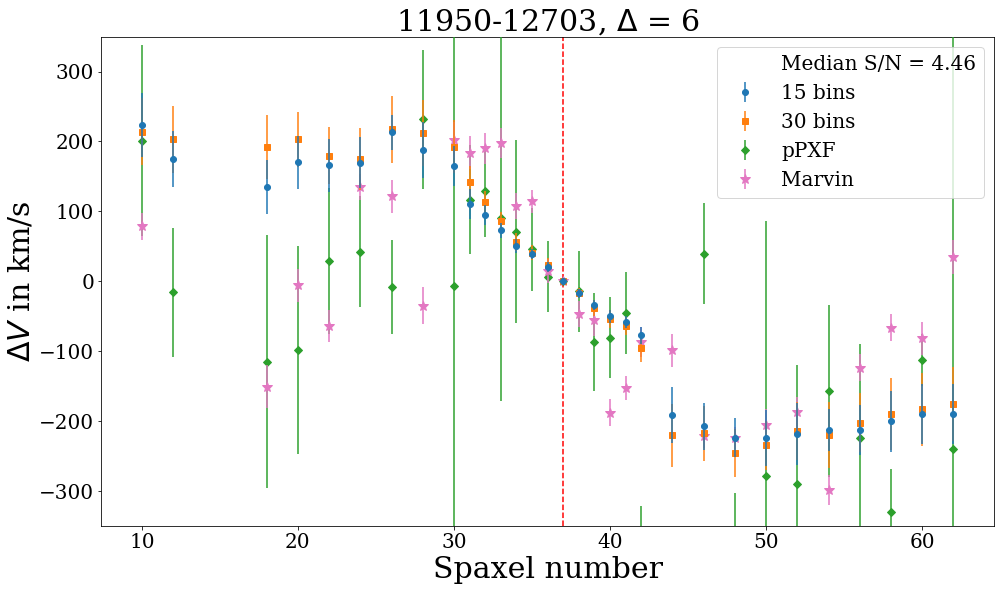}}
    \label{fig:b2}

\subfloat[Galaxy \#3]{\includegraphics[height=0.22\linewidth]{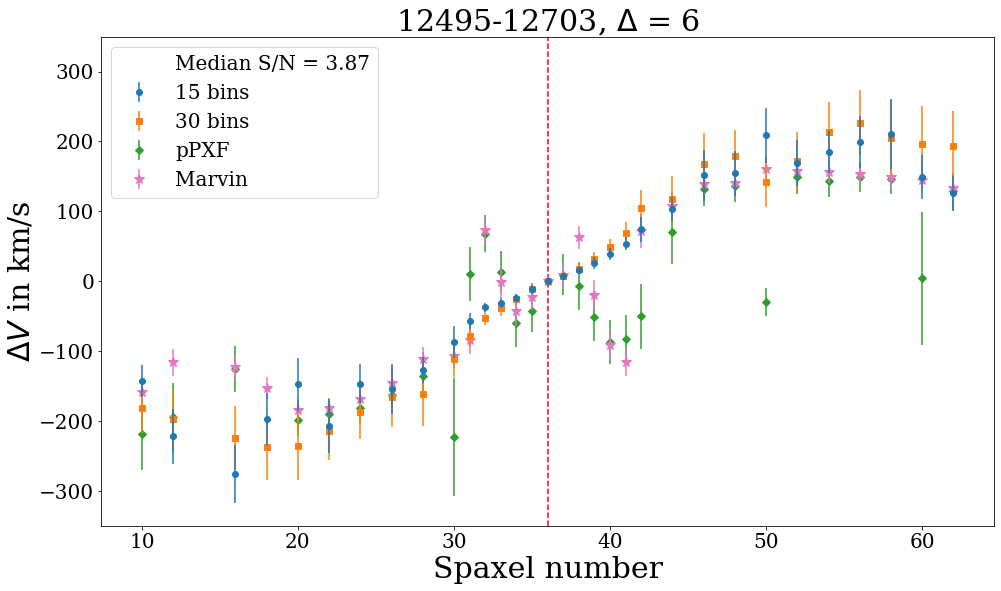}}
    \label{fig:c2}
\hspace{1em}
\subfloat[Galaxy \#4]{\includegraphics[height=0.22\linewidth]{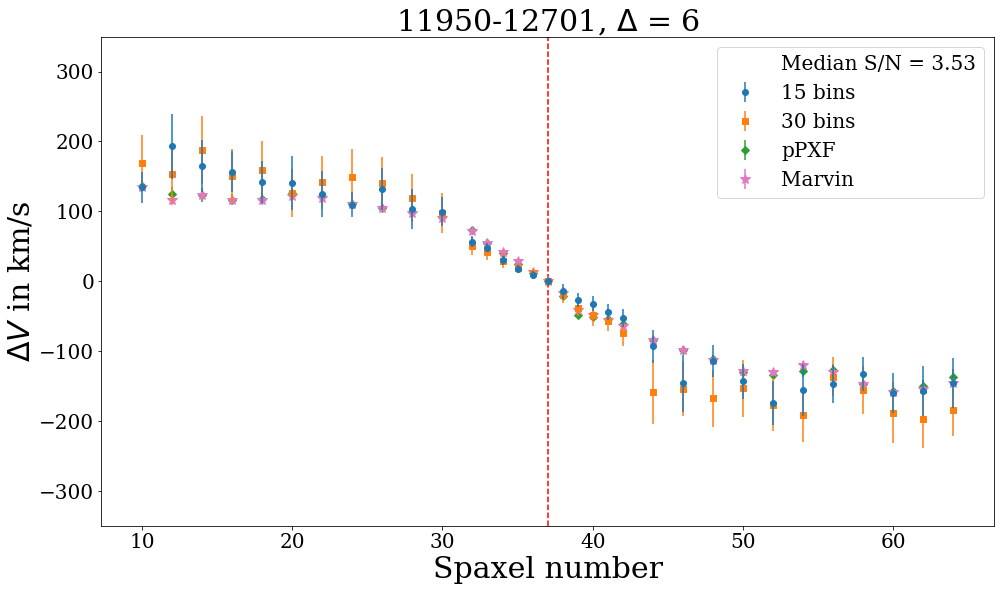}}
    \label{fig:d2}

\subfloat[Galaxy \#5]{\includegraphics[height=0.22\linewidth]{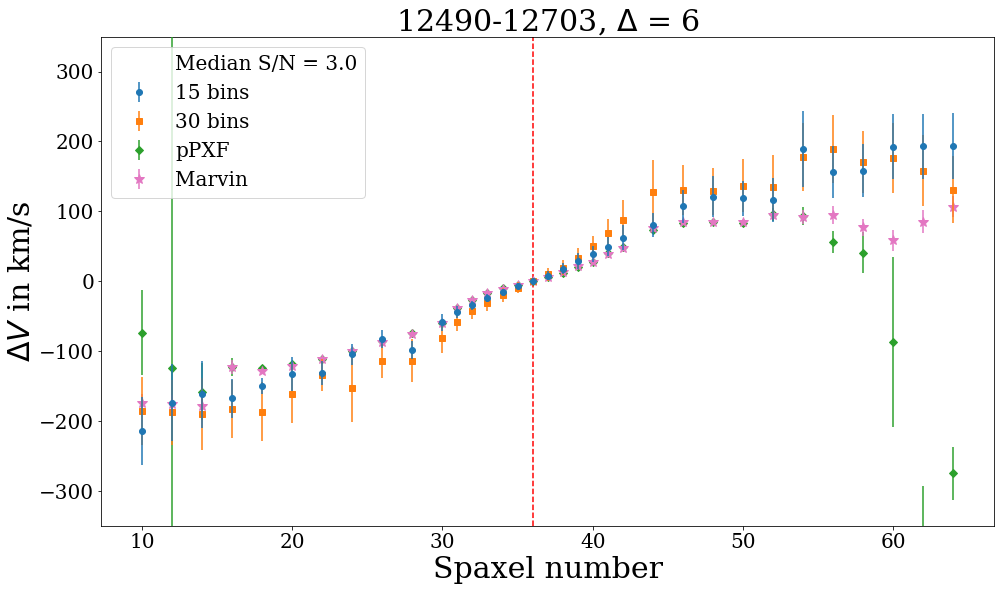}}
    \label{fig:e2}
\hspace{1em}
\subfloat[Galaxy \#6]{\includegraphics[height=0.22\linewidth]{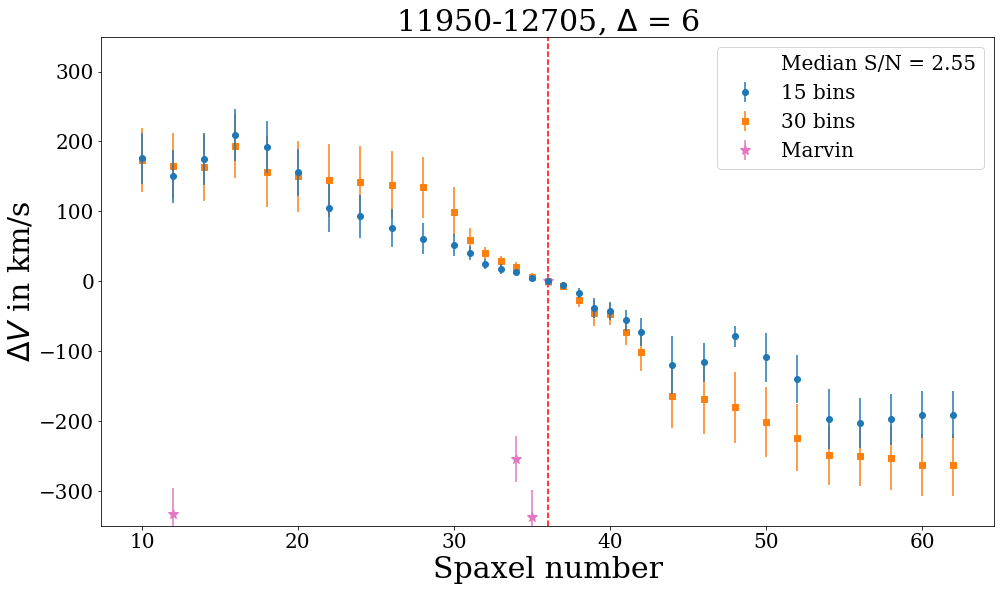}}
    \label{fig:f2}

\subfloat[Galaxy \#7]{\includegraphics[height=0.22\linewidth]{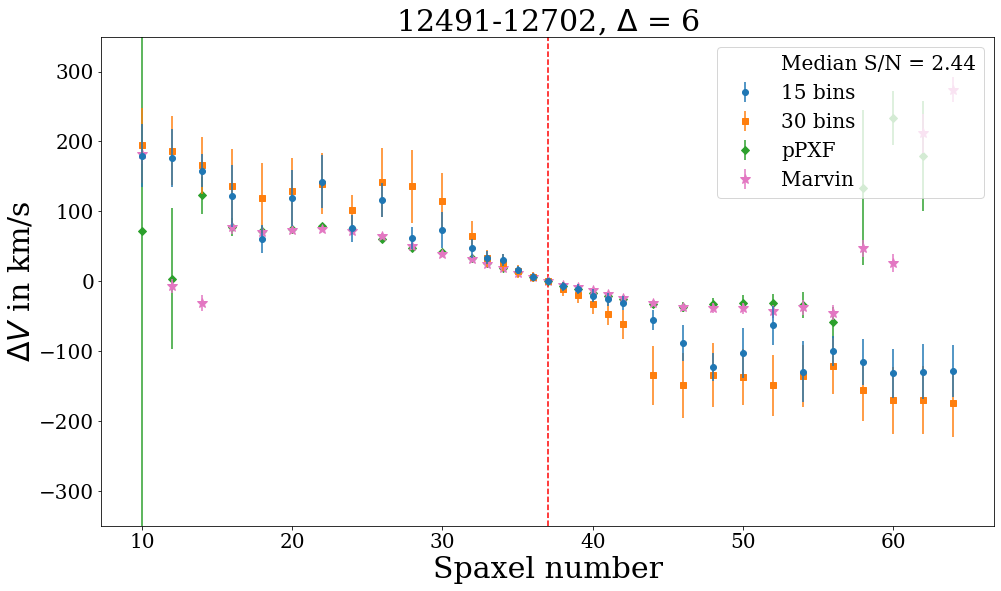}}
    \label{fig:g2}
\hspace{1em}
\subfloat[Galaxy \#8]{\includegraphics[height=0.22\linewidth]{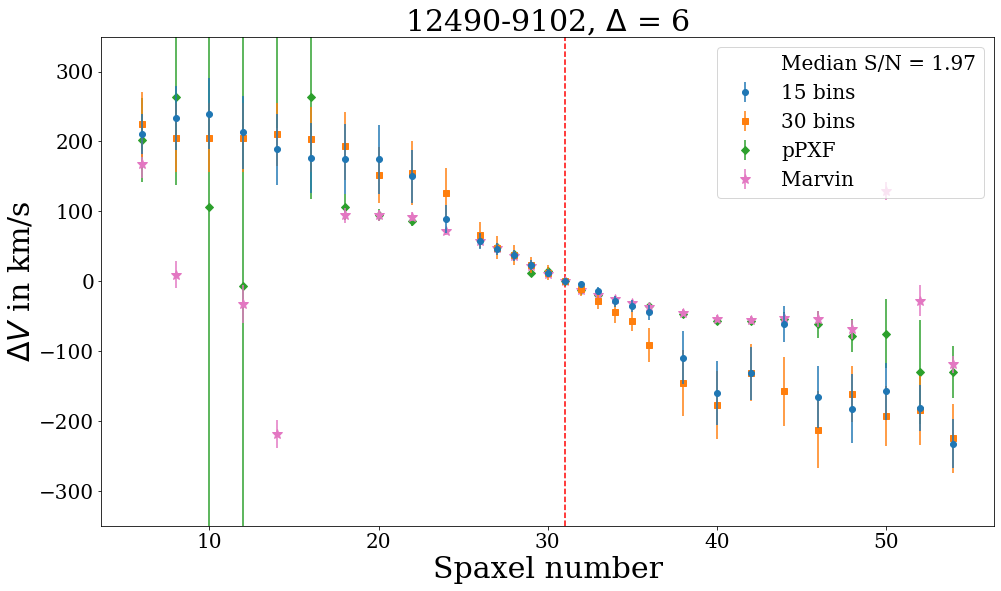}}
    \label{fig:h2}

\subfloat[Galaxy \#9]{\includegraphics[height=0.22\linewidth]{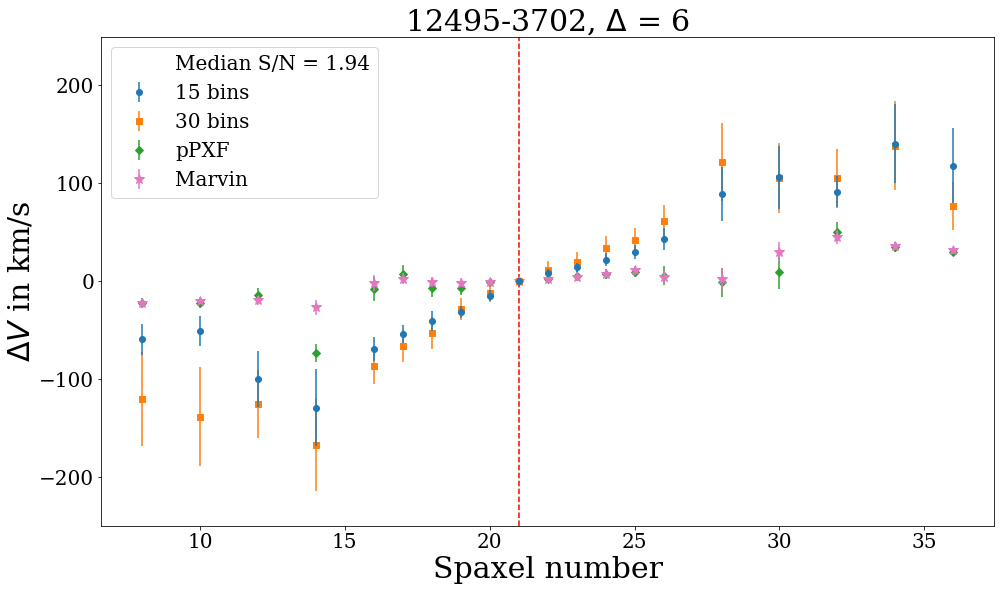}}
    \label{fig:i2}
\hspace{1em}
\subfloat[Galaxy \#10]{\includegraphics[height=0.22\linewidth]{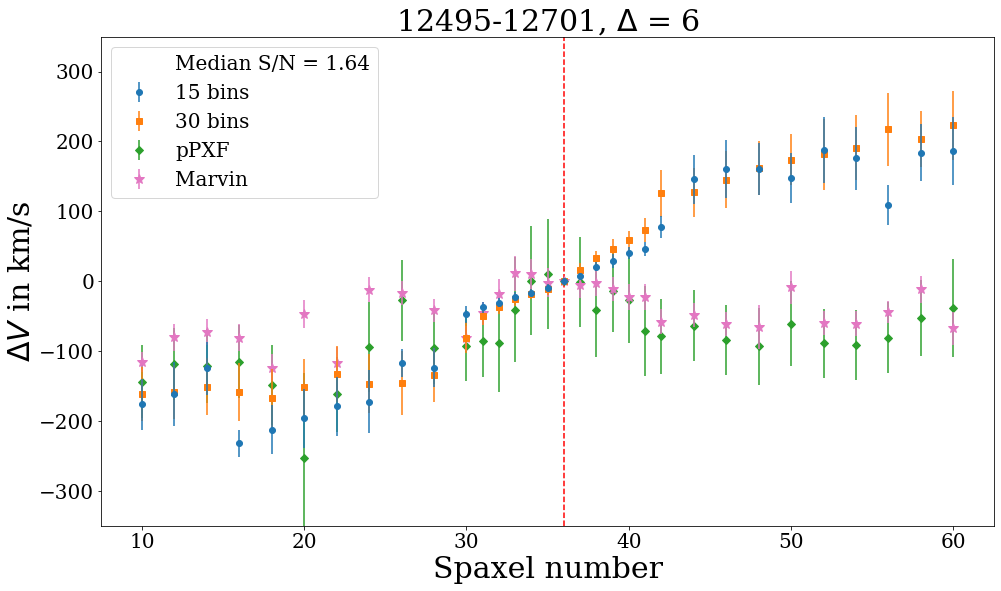}}
    \label{fig:j2}

\caption{\label{fig:fig4_2} 
Same as Figure~\ref{fig:fig4}, but using H$\alpha$ gas velocity estimation from pPXF and Marvin. Both pPXF and Marvin estimates differ from the stellar velocity estimation, depending on the characteristics of galaxies. Note that in galaxies \#1 and \#4 where the quality of data is better, we have impressive consistency between all estimations. 
}

\end{figure*}

\begin{figure}

\centering

\includegraphics[width = 0.8\linewidth]{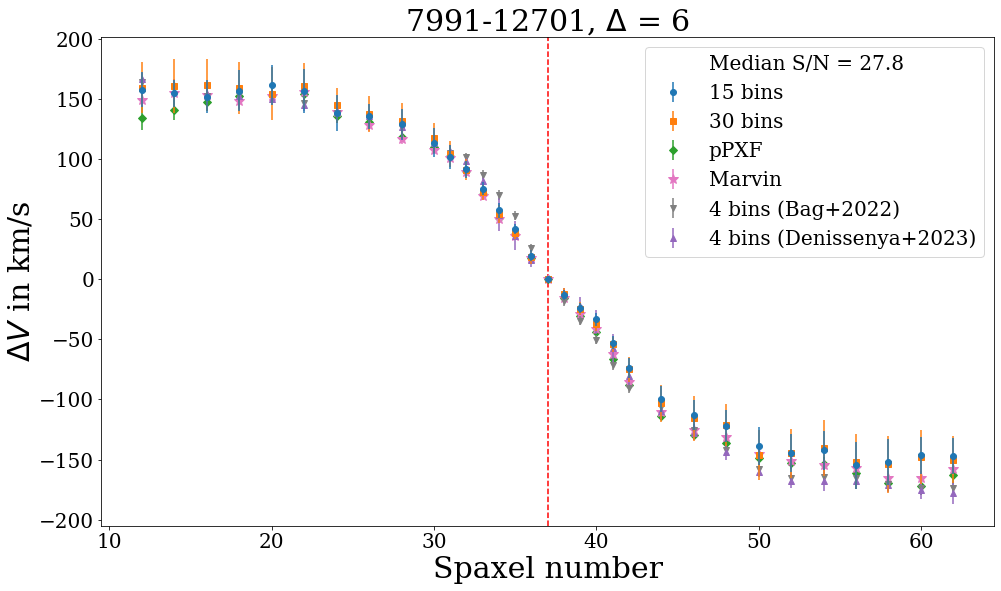}
\caption{\label{fig:fig5} 
As in Figure~\ref{fig:fig4}, but for the high $S/N$ galaxy $\#0$. In this case, we also plot the previous results from \cite{Bag:2021xym, Denissenya:2023tzw} (\textit{gray} and \textit{purple}). In this high $S/N$ regime, the different approaches match quite well.}  
  
\end{figure}

\section{Rotation Curve Results}
\label{sec:results}

Figure~\ref{fig:fig4} and~\ref{fig:fig4_2} present our derived velocities for the 10 low $S/N$ galaxies with those obtained from stellar and gas velocities in both pPXF and Marvin for the comparison, respectively. They show the final LOS velocities of the spaxels with respect to the centers of the respective galaxies. Recall that the median $S/N$ ranges from 5.55 for galaxy $\#1$ to 1.64 for $\#10$. Even for the lowest median $S/N$ case, we obtain reasonable-looking rotation curves and good consistency between the 15 and 30 bins analyzes. 

Both estimations from pPXF and Marvin involve two different methods of velocity estimation: stellar velocity and gas velocity (H$\alpha$ emission line of Balmer series). The dark matter distribution dominates the mass within galaxies, hence primarily driving the kinematics of the baryons within it. Hence, the observed kinematical profile is an essential tool in astronomy to model the dark matter profile of a galaxy. Both estimations have different characteristics, and the connection between stellar and gas velocity and the mass profile of the galaxy depends on the galaxy type. For example, there exists a well-defined relation between stellar velocity dispersion and mass, but this is limited to quiescent galaxies because these galaxies are dispersion dominated. On the other hand, some galaxies could have strong outflows from their central supermassive black hole, starbursts, or supernova explosions, which would influence the gas more significantly than the stars due to its fluid nature. In this paper, we will assume a simplistic model of galaxy dynamics ignoring the effects of these above-mentioned disturbances. This is apt for our work, as the purpose of this paper is to restricted to study of observed galaxy dynamics and not delve deeply into the influence of dark matter.


pPXF uses spectral features, such as stellar absorption and gas emission, to estimate the dynamics, as well as the spectral shape-profile to estimate the morphology of a galaxy. Upon examining pPXF results, they appear to perform well in estimating dynamics near the center for galaxies with the highest median $S/N$ galaxies. However, for lower $S/N$ galaxies, the dynamics derived from pPXF tend to have significantly higher uncertainty. Therefore, the pPXF results have diminished utility for low $S/N$ data. Interestingly, the effectiveness of different estimation methods varies depending on the characteristics of galaxies. For instance, galaxies \#2 and \#3 exhibit good results for stellar velocity in Figure~\ref{fig:fig4}, while galaxies \#1, \#4, and \#5 show better performance for gas velocity in Figure~\ref{fig:fig4_2}. Further discussion on this topic will be provided in Section~\ref{sec:ppxf}.

On the contrary, Marvin results are unfortunately even less reliable than pPXF in these situations in general. It appears that Marvin results are more reliable for gas velocity rather than stellar velocity, with unusually small uncertainty. From Figure~\ref{fig:fig4}, Marvin fails to construct rotation curves altogether for stellar velocity in most cases, providing only a single or a few values, typically calculated only at the center of the galaxy. These rotation curves are often flat, indicating either a lack of meaningful data or a failed fit. In contrast, for gas velocity in Figure~\ref{fig:fig4_2}, Marvin demonstrates better overall results, especially for galaxies \#1, \#4, and \#5. They also exhibit similar trends to those from pPXF. Consequently, Marvin results are generally considered highly unreliable for low $S/N$ galaxies for stellar velocity. They demonstrate better rotation curves for gas velocity compared to stellar velocity. However, similar to pPXF, Marvin also can only estimate velocities for relatively high $S/N$ galaxies, not for low $S/N$ galaxies.

As a crosscheck, Figure~\ref{fig:fig5} presents the rotation curve for the high $S/N$ galaxy $\#0$, calculated under the same conditions as Figure~\ref{fig:fig4}. Additionally, we include results for stellar velocity from both pPXF and Marvin, and the rotation curves estimated from ~\cite{Bag:2021xym, Denissenya:2023tzw} for comparison. Note our results are calculated from 15 bins and 30 bins with smoothing scale $\Delta = 6.0$ \AA{}, while results from ~\cite{Bag:2021xym, Denissenya:2023tzw} were estimated from 4 bins with $\Delta = 1.5$ \AA{}. 

In this case, all results, including those from pPXF and Marvin, appear consistent within uncertainty. Although estimates from our larger $\Delta$ value may lack precision compared to the smaller $\Delta$ used previously, the consistency remains noteworthy. And of course, for the low $S/N$ galaxies we find that the larger $\Delta$ proves essential to tame bias from the noise. 
Overall the iterative smoothing plus cross-correlation techniques are found to be effective across a wide range from high $S/N$ galaxies down to quite low $S/N$ galaxies.

\section{Comparision with Penalized PiXel-Fitting (pPXF)}
\label{sec:ppxf}

In this section, we focus on comparing our estimation method with pPXF. These comparisons reveal intriguing insights, particularly concerning the differences between the estimations derived from stellar velocity and gas velocity. Unlike pPXF, Marvin estimations lack strong constraining power for stellar velocity estimation. However, pPXF generally produces better results for both stellar and gas velocities compared to Marvin, especially for higher $S/N$ galaxies.


pPXF is a SED fitting code that uses maximum penalized likelihood in pixel space to solve for the line-of-sight velocity distribution (LOSVD) described by a Gauss-Hermite series of stars within either a stellar population or a galaxy spectra, including photometry. If required, pPXF can also solve for likely multiple stellar populations, their morphology, and can perform additional multi-component emission line fitting to study gas emission. pPXF takes stellar population models from a standard library, such as MILES~\cite{2010MNRAS.404.1639V, MILES} and uses it to find the best solution with multiple stellar populations and morphology. Due to its simplicity and ease of use, in the last few decades, pPXF has largely become the standard SED fitting code for fitting stellar populations and galaxy spectra in the optical. For our work, we have specifically fit one stellar component along with one gas component for each of our spaxels within the galaxy using pPXF.

In \cite{Bag:2021xym}, they compared their estimated velocity reconstructions from different noise sets with the standard traditional fitting approach based on the pPXF, using MILES templates. \cite{Bag:2021xym} showed that pPXF result matched well for the low noise set, however, it does not fit well for a high noise set, with larger errors for low $S/N$ cases. In general, $S/N$ of the galaxy is the highest around the center and it decreases gradually for the outskirt of the galaxy as shown in Figure~\ref{fig:fig_SN}. Although the pPXF method matched well in the central region, it failed to accurately reflect the rotation curve in the outskirts of the galaxy, corresponding to the low $S/N$ region.

In Figures~\ref{fig:fig4} and~\ref{fig:fig4_2}, we compare our estimations with pPXF results for stellar and gas velocities respectively, as we focus on the analysis of low $S/N$ galaxies. As previously shown by~\cite{Bag:2021xym}, pPXF is not accurate for low $S/N$ cases. \\

We should note here that in the event of a possible misalignment between the stellar and gas components in the raw spectra, we would detect such cases in our analysis if we separate the gas and stellar components relatively. In Appendix~\ref{sec:appendix_b} we 
 mask the gas emission lines and redo our analysis for few galaxies of our sample (with higher signal to noise ratios) to test and demonstrate such consistencies. The galaxies in our sample show no such misalignment but this remain an interesting topic for our future works to study galaxies with misalignment between their stellar and gas components using our approach.\\

As discussed in Section~\ref{sec:results}, the accuracy of estimation varies depending on the specific characteristics of galaxies and their median $S/N$ values. Generally, higher $S/N$ galaxies tend to exhibit closer agreements between our estimation and those obtained from pPXF. For example, galaxies \#2 (median $S/N$ = 4.46) and \#3 (median $S/N$ = 3.87) in Figure~\ref{fig:fig4} show a notable alignment between our results and pPXF estimated from stellar velocity, particularly in the central regions where the $S/N$ is high. However, as we move towards the outskirts of galaxies, discrepancies become more prominent, accompanied by increased uncertainties. Galaxies \#1 and \#4 also demonstrate good matches with our estimations. Conversely, for galaxies with lower $S/N$ values, such as those below galaxy \#5 (median $S/N$ = 3.00), pPXF tends to perform poorly, yielding unrealistic larger uncertainties when it goes to the outskirts. In short, pPXF estimations for stellar velocity start to improve significantly for $S/N \gtrsim 3.00$.

On the other hand, pPXF estimations derived from gas velocity in Figure~\ref{fig:fig4_2} illustrate better performance compared to those calculated from stellar velocity overall. Notably, galaxies \#1 (median $S/N$ = 5.55) and \#4 (median $S/N$ = 3.53) show remarkable agreement across all estimations. Galaxies \#2 and \#3, which gave satisfactory results in stellar velocity, have numerous fluctuations in gas velocity estimations. For galaxy \#5 onward, such as galaxies \#5, \#7, and \#8, accurate estimations are primarily confined to the central spaxels, with pPXF results deteriorating notably towards the edges. Moreover, for galaxies with even lower $S/N$ values, such as galaxies \#9 and \#10, pPXF estimations deviate significantly from the expected patterns, indicating a loss of accuracy.\\

It is important to note that measuring the uncertainties in our approach fundamentally differs from that of pPXF. While pPXF derives its error estimates from the original uncorrelated spectrum (template), our method calculates the final uncertainties from HMC based on the derived mean and standard deviation of different $\Delta V$s from the cross-correlation of multiple anchor spaxels with respect to other spaxels in the data. This approach is particularly beneficial in complex datasets, as it allows us to account for variations across different measurements, leading to potentially more consistent estimates, especially in low $S/N$ scenarios. However, pPXF’s method of fitting a single spectrum will result in higher statistical errors in the fainter, outer regions where the spaxels are expected to have a low $S/N$.\\

In summary, the two different estimations from pPXF for stellar and gas velocities offer distinct perspectives. Some galaxies exhibit better performance for stellar velocity, whereas others show better alignment with gas velocity due to the dependence on spectral lines. For relatively higher $S/N$ galaxies, pPXF could estimate reasonable results, such as in galaxies \#1, \#2, \#3, and \#4. Note that almost perfect agreement for galaxies \#1 and \#4 with gas velocity, providing strong evidence supporting the consistency between our estimations and pPXF. Despite their relatively higher $S/N$ values, it is clear that they still qualify under the category of low $S/N$ galaxies.

Indeed, our estimation method demonstrates the capability to estimate accurate galaxy rotation curves regardless of the characteristics of galaxies. For example, while galaxies \#2 and \#3 perform better for stellar velocity, galaxies \#1 and \#4 show superior alignment with gas velocity. Our estimations maintain consistency without assuming any specific velocity profile, offering improved precision as well.


From another perspective, there appears to be consistency between our estimates and pPXF for the central region around the anchor spaxel in some galaxies. However, pPXF fails to accurately reconstruct rotation curves beyond certain boundaries. Specifically, some of the results at the edges have excessively large error bars. This suggests that while pPXF can estimate rotation curves around the center of galaxies, it struggles with accurately capturing dynamics beyond certain regions.

In other words, the uncertainty of the pPXF result at the outskirts of galaxies is significantly larger than ours, sometimes leading to unrealistically large errors. It appears that pPXF results are unable to accurately estimate the entire rotation curves for low $S/N$ galaxies, especially at the edges of each galaxy. In particular, pPXF results fail to reconstruct rotation curves when the median $S/N$ is lower, as observed in galaxies $\#7$, $\#8$, $\#9$, and $\#10$.

Additionally, considering the consistent results of Figure~\ref{fig:fig5} and agreement in estimations from relatively higher $S/N$ galaxies such as galaxies \#1, \#2, \#3, and \#4, we have shown that our estimated velocities remain consistent regardless of the smoothing scale $\Delta$ for higher $S/N$ galaxies. Moreover, our estimation shows robust performance even for lower $S/N$ galaxies. Therefore, we can expect our velocity estimation from Figure~\ref{fig:fig5} to be reliable even for lower $S/N$ galaxies in Figures~\ref{fig:fig4} and~\ref{fig:fig4_2}.

We emphasize that what we demonstrate in our paper is not only about the better precision in reconstruction of rotation curves, but the ability of our method to derive the rotation curve for low $S/N$ galaxies while pPXF or Marvin may fail to determine the shape of the rotation curve in some of these cases, such as galaxies \#7, \#8, \#9, and \#10.


Consequently, our estimations outperform those from pPXF in all low $S/N$ galaxies, exhibiting consistent small uncertainty regardless of $S/N$.

\section{Conclusions and Discussion}
\label{sec:conclusion}

Internal dynamics of galaxies, including the rotation curves, provide valuable insights into revealing the properties of dark matter. They can be utilized not only for investigating dark matter but also for creating velocity maps in the development of kinematic lensing techniques for upcoming surveys aimed at unveiling unsolved cosmological questions~\cite{DiGiorgio:2021egh, S:2022htv}. 
Traditional methods for estimating rotational velocity typically involve fitting a template and calculating the velocity based on the Doppler shift of various spaxels.

In this paper, we explore a template-free approach for estimating the velocity field in galaxies using iterative smoothing and cross-correlation techniques applied to the spectra of IFU data. We demonstrate the effectiveness of this method by applying it to 10 very low $S/N$ MaNGA galaxies, illustrating its capability to produce accurate velocity estimations in such challenging conditions.

By comparing one spaxel to another, we can obtain LOS wavelength shifts resulting from the Doppler effect caused by galaxy rotation. Iterative smoothing of spectra is applied before cross-correlation, which is crucial, especially for low $S/N$ galaxies since raw spectra are often contaminated by strong uncertainties that can obscure the true wavelength shifts. 

Cross-correlation is used to calculate velocity differences $\Delta V$. We find that subdividing the spectra into 15 or 30 wavelength bins is crucial for estimation with low $S/N$ galaxies, since the spectra between pairs of spaxels often have different shapes due to stellar population shifts, with high noise complicating its recognition. 

In our technique all of these different aspects are important: proper smoothing scales $\Delta$, the number of bins for cross-correlation, and the consistency criteria, to enable robustness for low $S/N$ galaxies. 


We carried out a direct comparison with the traditional fitting method of Penalized PiXel-Fitting (pPXF), as well as the results provided by Marvin, using two different velocity estimations: stellar velocity and gas velocity. Our velocity estimation method significantly outperforms both pPXF and Marvin regardless of velocity profiles, particularly at low $S/N$. pPXF struggles to calculate rotation curves accurately for low $S/N$ galaxies, often resulting in near-zero velocities or unrealistically large error bars. In contrast, our estimation method consistently provides reliable results across all 10 galaxies analyzed, passing several crosschecks, and demonstrating smaller uncertainty.

Indeed, an advantage of our approach is its independence from any templates and specific velocity profiles, as well as its lack of reliance on strong spectral features. We do not need to consider the specific characteristics of galaxies, which can vary widely and pose challenges for accurate estimation, particularly in cases of low $S/N$ galaxies. This independence allows for more robust and accurate estimations, making our method particularly effective in scenarios where traditional methods may struggle due to low $S/N$.


In conclusion, our model independent approach to galaxy velocity estimation, utilizing iterative smoothing and cross-correlation methods, appears to be a clear improvement, particularly for galaxies with low $S/N$. We have demonstrated the effectiveness of our method across a range of $S/N$ levels, down to median $S/N=1.64$, providing accurate estimations with smaller uncertainty. 
Moving forward, we plan to extend our 
approach to merge with that of \cite{Denissenya:2023tzw} that goes from 1D to 2D, for reconstructing full 2D galaxy velocity maps in low $S/N$ galaxies.

\section*{Acknowledgements}

The authors would like to thank Yongmin Yoon, Wuhyun Sohn and Ho Seong Hwang for useful discussions. This work was supported by the high performance computing cluster Seondeok at the Korea Astronomy and Space Science Institute (KASI). A.~S. would like to acknowledge the support by National Research Foundation of Korea NRF-2021M3F7A1082056 and the support of the Korea Institute for Advanced Study (KIAS) grant funded by the government of Korea. S.~B. acknowledges the funding provided by the Alexander von Humboldt Foundation. E.~L. is supported in part by the U.S.\ Department of Energy, Office of Science, Office of High Energy Physics, under contract no.\ DE-AC02-05CH11231.

 \appendix
 \section{Comparison of Different Smoothing Scale $\Delta$}
 \label{sec:appendix}

Estimating galaxy rotation curves for low $S/N$ galaxies is challenging, particularly due to their sensitivity to the smoothing scale $\Delta$. As discussed in Section~\ref{sec:M3}, a smaller $\Delta = 1.5$ \AA{} for iterative smoothing is adequate for galaxy \#0 (median $S/N$ = 27.8). However, this approach is not suitable for low $S/N$ galaxies.

\begin{figure}

\centering

\subfloat[15 bins]{\includegraphics[width = 0.48\linewidth]{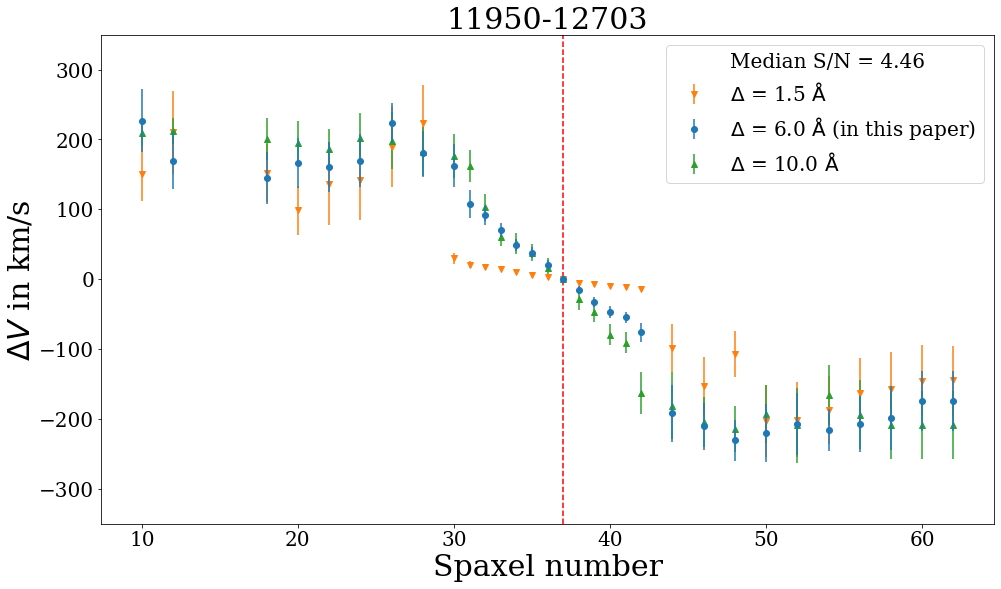}}
    \label{fig:15bins_app1}
\subfloat[30 bins]{\includegraphics[width = 0.48\linewidth]{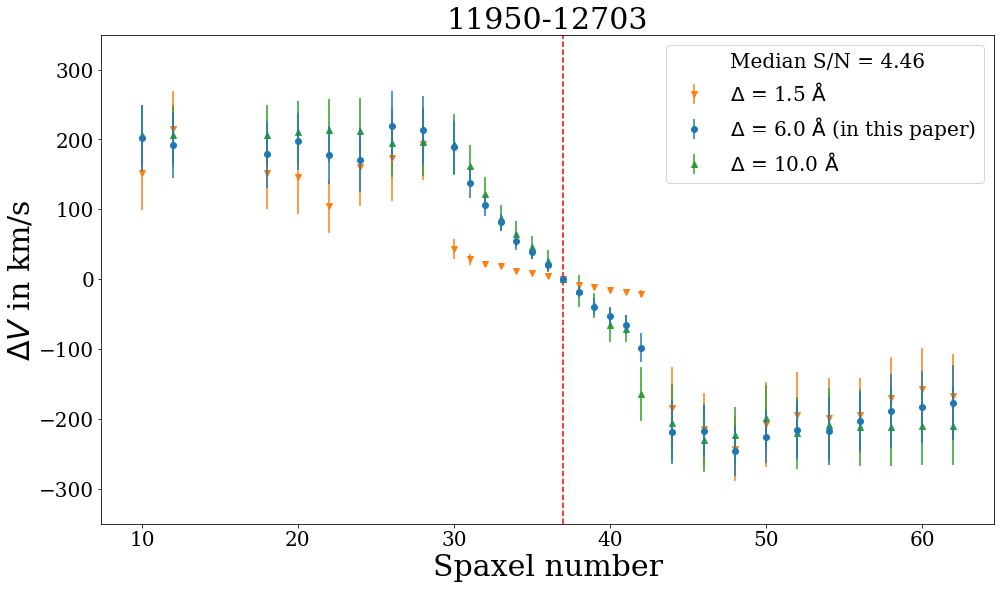}}
    \label{fig:30bins_app1}

\caption{\label{fig:fig_app1} The galaxy rotation curves for galaxy \#2 are compared using different smoothing scales $\Delta$ = 1.5, 6.0, and 10.0 \AA{}, with the spectra range divided into 15 bins (\textit{left}) and 30 bins (\textit{right}). We select $\Delta$ = 6.0 \AA{} as the main result of this paper. Note that there are discontinuous regions around anchor spaxels in $\Delta$ = 1.5 \AA{}.}  
  
\end{figure}

Figure~\ref{fig:fig_app1} illustrates the comparison of galaxy rotation curves for galaxy \#2 using three different $\Delta$ values: $\Delta$ = 1.5, 6.0, and 10.0 \AA{}. For smaller $\Delta$, such as $\Delta$ = 1.5 \AA{}, we see overly small standard deviations and some artificial linear regions around the anchor spaxels, regardless of the number of bins. When using smaller $\Delta$ values, cross-correlation performs well in the higher $S/N$ region, particularly around the anchor spaxels. However, beyond certain boundaries, it produces a correlation coefficient curve with excessive fluctuations and peaks, leading to significant discontinuities despite dividing them into numerous bins.

\begin{figure}

\centering

\subfloat[15 bins]{\includegraphics[width = 0.48\linewidth]{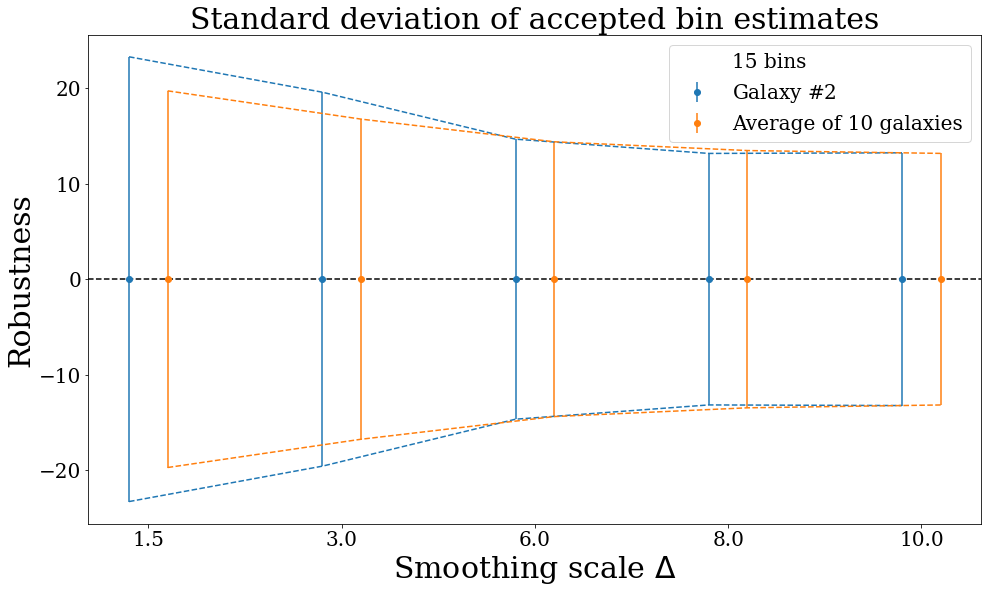}}
    \label{fig:15bins_app2}
\subfloat[30 bins]{\includegraphics[width = 0.48\linewidth]{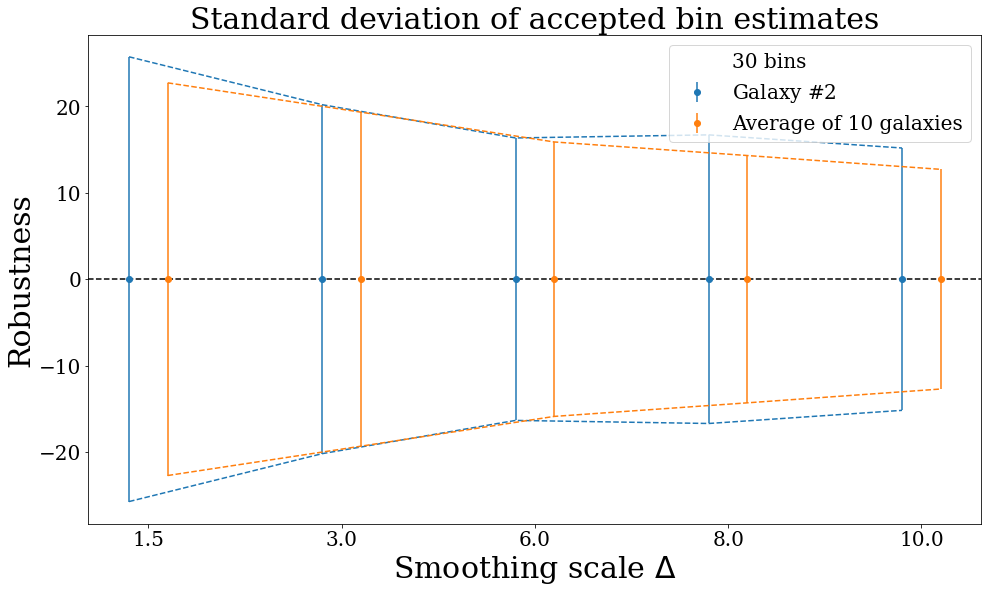}}
    \label{fig:30bins_app2}

\caption{\label{fig:fig_app2} The standard deviation of accepted bin estimates for galaxy \#2 (\textit{blue}) and the average of 10 low $S/N$ galaxies (\textit{orange}), as a function of $\Delta$ for 15 bins (\textit{left}) and 30 bins (\textit{right}). Note that the scatters are the greatest for $\Delta$ = 1.5 \AA{}, and they gradually decrease until reaching $\Delta$ = 6.0 \AA{}. When $\Delta$ $\ge$ 6.0 \AA{}, the standard deviations saturate, showing minimal change.
}  

\end{figure}

Interestingly, these discontinuities disappear when $\Delta$ $\ge$ 6.0 \AA{}. Therefore, we evaluate the robustness of estimation for five different values: $\Delta$ = 1.5, 3.0, 6.0, 8.0, and 10.0 \AA{}. Figure~\ref{fig:fig_app2} demonstrates the robustness of estimation as a function of $\Delta$ for both galaxy \#2 and the average of 10 low $S/N$ galaxies. It represents the scatter of the standard deviation, i.e., the standard deviation of the standard deviation. These scatters are largest for $\Delta$ = 1.5 \AA{}, and they gradually decrease until $\Delta$ = 6.0 \AA{}. Beyond $\Delta$ = 6.0 \AA{}, they saturate and show almost no change. This suggests that the scatter of the standard deviation saturates after a certain $\Delta$ for low $S/N$ galaxies, supporting the robustness of the estimation when $\Delta$ $\ge$ 6.0 \AA{}. Therefore, we choose $\Delta$ = 6.0 \AA{} for the optimal result.

 \section{Comparison with Masking Gas Emission Lines}
 \label{sec:appendix_b}

\begin{table}
\caption{\label{tab:table2} List of emission lines used for masking spectra.} 
\,\\ 
\centering
\begin{tabular}{|c|c|c|c|c|c|}
\hline
Line & Wavelength (\AA{}) & Line & Wavelength (\AA{}) \\
\hline

O II & 3727 & N II & 5755  \\
Ne III & 3868 & He I & 5876  \\
H $\epsilon$ & 3970 & Fe VII & 6087  \\
H $\delta$ & 4101 & O I & 6300  \\
H $\gamma$ & 4340 & S III & 6312  \\
O III & 4363 & O I & 6363  \\
He II & 4686 & Fe X & 6375 \\
H $\beta$ & 4861 & N II & 6548  \\
O III & 4959 & H $\alpha$ & 6563 \\
O III & 5007 & N II & 6583  \\
N I & 5199 & S II & 6716 \\
He II & 5411 & S II & 6731  \\
O I & 5577 & Ar III & 7136 \\
\hline
\end{tabular}
    
\end{table}
 
\begin{figure}

\centering

\subfloat[Galaxy \#1]{\includegraphics[width = 0.48\linewidth]{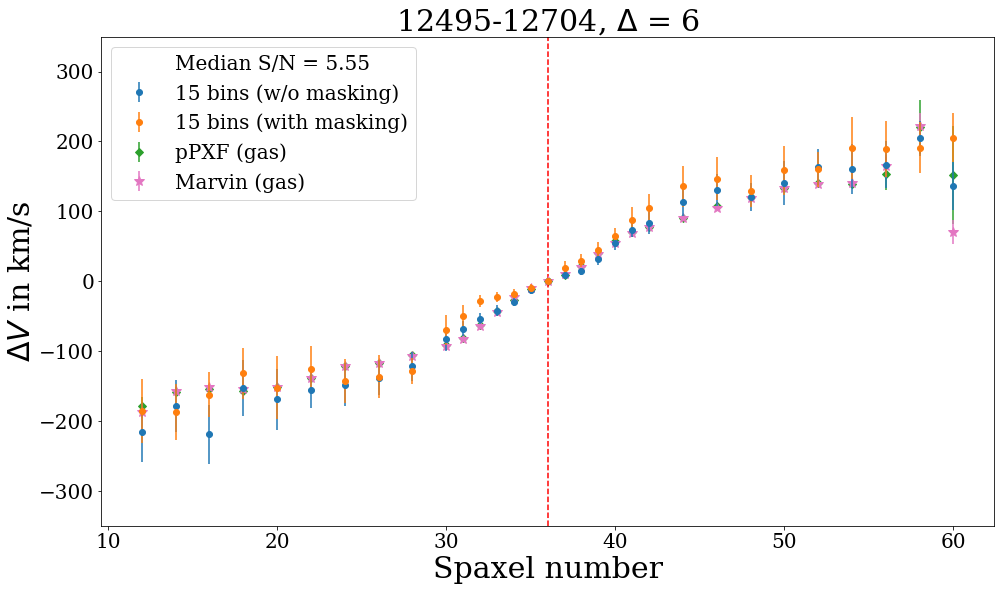}}
    \label{fig:gal1}
\subfloat[Galaxy \#2]{\includegraphics[width = 0.48\linewidth]{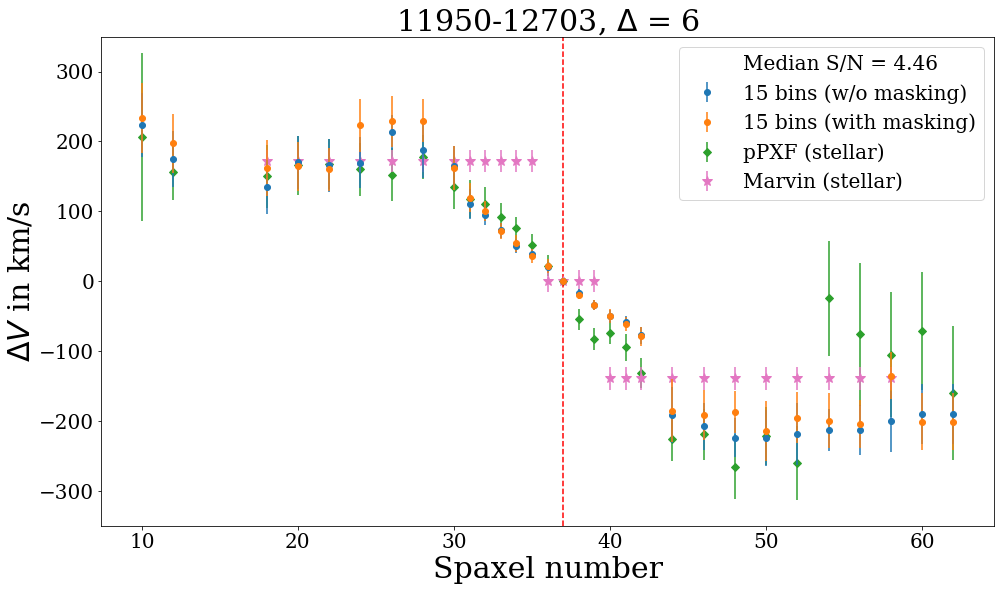}}
    \label{fig:gal2}
\subfloat[Galaxy \#3]{\includegraphics[width = 0.48\linewidth]{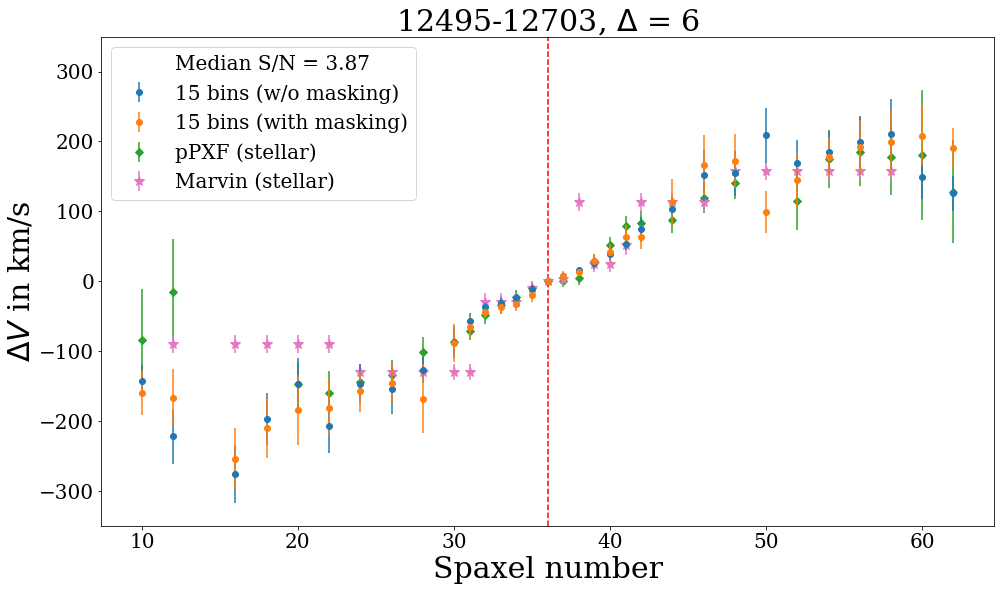}}
    \label{fig:gal3}
\subfloat[Galaxy \#4]{\includegraphics[width = 0.48\linewidth]{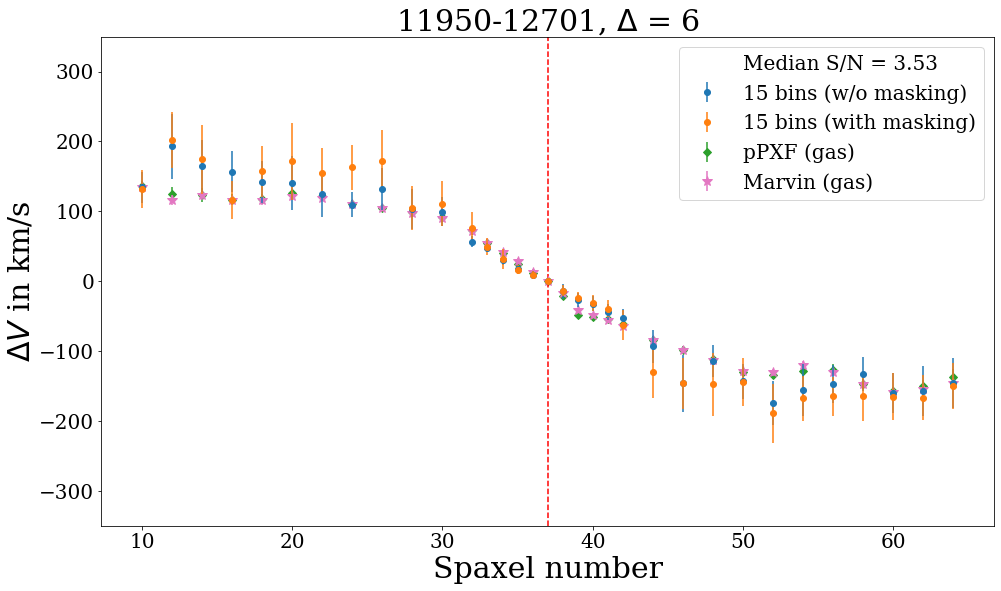}}
    \label{fig:gal4}

\caption{\label{fig:fig_mask} The galaxy rotation curves for galaxies \#1, \#2, \#3, and \#4 are presented both with (\textit{orange}) and without (\textit{blue}) masking specific gas emission lines. Additionally, the results from pPXF (\textit{green}) and Marvin (\textit{pink}) for gas or stellar analysis, which are considered best for each galaxy respectively, are also plotted. }  
  
\end{figure}

In our study, we utilize the full spectrum, incorporating both stellar absorption and gas emission lines, resulting in a combined measurement of stellar and gas velocities. Our rotation curves are derived from incorporating the entire spectral data. When compared to the Penalized Pixel-Fitting (pPXF) method, which separates these velocities, we generally observe consistency with the component that pPXF captures most effectively. For instance, in galaxies \#1 and \#4, our results align well with the gas velocities from pPXF (for these galaxies, pPXF provides better results with gas velocities than with stellar velocities). Meanwhile, in galaxies \#2 and \#3, our results align well with the stellar velocities from pPXF (for these galaxies, pPXF provides better results with stellar velocities than with gas velocities).

Among the galaxies we analyzed, there are no instances where the stellar and gas velocities from pPXF show contradictory results (due to misalignment of the stellar and gas velocity fields). Indeed, at the spectral resolution, R$\sim$2000 of SDSS-MaNGA, it may not always be possible to disentangle the misalignment between the kinematics of gas and stars within a distant galaxy. However, if such a case exists, considering the quality of the data, our method can potentially detect such cases by masking the gas emission lines and comparing the results. While this could be an interesting case of study for our future works to focus on such types of galaxies, to demonstrate this approach, we have masked the emission lines for our first four galaxies (\#1, \#2, \#3, and \#4) and compared the results of our method.

Table~\ref{tab:table2} provides the list of emission lines used for masking, obtained from the SDSS Science Archive Server (SAS), while Figure~\ref{fig:fig_mask} shows the rotation curves for these galaxies, presented both with and without masking the gas emission lines, as listed in Table~\ref{tab:table2}. Additionally, the better results from pPXF and Marvin for gas and stellar analyses are included for each plot. The results demonstrate clear consistency between the two cases, whether the emission lines are masked or not. While there are obviously slight differences in the results, the overall shape of the rotation curves has remained unchanged and consistent with each other.

This suggests that our method does not solely rely on gas emission lines and the overall properties of the data are responsible for the derived rotation curves. This is particularly evident when examining panels (b) and (c) of Figure \ref{fig:fig_mask}, where the results of our approach (with and without masking) are compared with pPXF and Marvin results with stellar velocities. We can see that even by masking the gas emission lines, our approach provides consistent results with pPXF in these galaxies. It is important to note that we selected these four galaxies for masking the gas emission lines and comparison since they have a relatively high $S/N$ in their spectra so we have some clear results from pPXF. In the galaxies with lower $S/N$, pPXF and Marvin may not be able to provide a sensible form for the rotation curve, as we demonstrated in the paper.


\bibliographystyle{JHEP}
\bibliography{biblio}


\end{document}